\documentclass[final,3p,12pt]{elsarticle}
\usepackage{amssymb}
\usepackage{amsmath}
\usepackage{amsthm}
\usepackage{tikz}
\usepackage{hyperref}
\usepackage{soul} 
\usepackage{microtype}
\journal{Finite Fields and Their Applications}

\newtheorem{proposition}{Proposition}
\newtheorem{lemma}{Lemma}
\newtheorem{theorem}{Theorem}
\newtheorem{corollary}{Corollary}
\theoremstyle{remark}
\newtheorem{remark}{Remark}
\theoremstyle{definition}

\newtheorem{problem}{Problem}

\newtheorem{example}{Example}
\newcommand{\FF}{\mathbb{F}}
\newcommand\param[5]{{(#1,#2)^{#3,#4}_{#5}}}
\newcommand\para[2]{{(#1,#2)}}
\def\Orb{\mathrm{Orb}}%
\def\qp{{q}} 
\newcommand{\alert}[2][magenta]%
{{\color{#1}\mbox{[\hspace{-0.4ex}[}#2\mbox{]\hspace{-0.4ex}]}}}

\begin{document}

\begin{frontmatter}
\title{Multispreads\footnote{This is the accepted version of the manuscript published in the Finite Fields and their Applications 108, 102675(1--25), 
\url{https://doi.org/10.1016/j.ffa.2025.102675}. \copyright 2025; available under the CC-BY-NC-ND 4.0 license. 
}}


\author[im]{Denis~S.~Krotov}
\ead{dk@ieee.org}
\author[im]{Ivan~Yu.~Mogilnykh} 
\ead{ivmog84@gmail.com}
\affiliation[im]{organization={Sobolev Institute of Mathematics},
            addressline={pr-kt Akademika Koptyuga 4}, 
            city={Novosibirsk},
            postcode={630090}, 
            country={Russia}\\[7mm]
\mbox{}\hfill{}{To the memory of Olof Heden}
}

\begin{abstract}
Additive one-weight codes over a finite field of non-prime order are equivalent to special subspace coverings of the points of a projective space, which we call multispreads. The current paper is devoted to the characterization of the parameters of multispreads, which is equivalent to the characterization of the parameters of additive one-weight codes and, via duality, of additive completely regular codes of covering radius~$1$ (intriguing sets). We characterize these parameters for the case of the prime-square order of the field and make a partial characterization for the prime-cube case and the case of the fourth degree of a prime, including a complete characterization for orders $8$, $27$, and $16$.
\end{abstract}



\begin{keyword}
spreads \sep multispreads \sep additive codes \sep one-weight codes \sep completely regular codes \sep intriguing sets

\MSC[2020] 51E23 \sep 94B05 \sep 51E22
\end{keyword}

\end{frontmatter}


\tableofcontents

\section{Introduction}

    By $\FF_q$, we denote the Galois field of order $q$;
    by $\FF_q^m$,
    the standard vector space of dimension $m$ over $\FF_q$;
    by $\FF_q^{m*}$,
    the set $\FF_q^m\setminus \{0\}$.
A $\param{\lambda}{\mu}{t}{m}{q}$-{\it multispread} \cite{Kro:mufold} is a multiset $S$ of subspaces of $\FF_q^m$ having dimensions at most~$t$ such that
\begin{equation}\label{eq:def-lm}
\lambda=\sum\limits_{U\in S}
(q^{t-\dim(U)}-1)
\quad\mbox{and}\
\mu=\sum\limits_{U\in S:\ x\in U}q^{t-\dim(U)},
\mbox{ for every $x\in\FF_q^{m*}$}
\end{equation}
or, equivalently, the following multiset equality takes place:
\begin{equation}\label{eq:def-mu}
    \lambda\times\{0\}
    +\mu\times \FF_q^{m*}
    =
    \sum_{U\in S}
    \big(q^{t-\dim(U)}\times U - \{0\}\big).
\end{equation}
For a $\param{\lambda}{\mu}{t}{m}{q}$-multispread,
we also use shorthand notations $\param{\lambda}{\mu}{t}{m}{}$-, $\para{\lambda}{\mu}$-{\it multispread} or simply {\it multispread}. The subspaces in a $\param{\lambda}{\mu}{t}{m}{q}$-multispread are allowed to have any dimension from~$0$ to~$t$, and we refer to~$t$ as the
{\it pseudodimension}
of any such subspace.

Multispreads are direct generalizations
of well-known spreads
($\param{0}{1}{t}{m}{}$-multispreads)
and multifold
spreads ($\param{0}{\mu}{t}{m}{}$-multispreads) in the following manner.
A spread ($\mu$-fold spread) is a partition (respectively, exact $\mu$-fold
covering) of the nonzero vectors of the space~$\FF_q^m$ into $t$-subspaces.
In such a partition, the all-zero vector is not considered because it is
covered as many times as the number of subspaces in the partition
(which is uniquely calculated from $q$, $m$, $t$, and $\mu$).
To generalize to multispreads, we replace ``subspace'' by the span of
arbitrary $t$ vectors, not necessarily linearly independent.
Considered as a multiset (in \cite{Kro:mufold}, it is called
a ``multisubspace''),
such a span~$V$ has the same cardinality~$q^t$,
regardless of the actual dimension $\dim(V)$,
and  every its element, including the all-zero vector, has multiplicity $q^{t-\dim(V)}$.
Now we see that a multispread is an exact $\mu$-fold covering of the nonzero vectors
of the space by such multisets.
Unlike for classical spreads, for multispreads the multiplicity of covering
of the zero is not
uniquely determined by the parameters,
and is denoted by
$\lambda+n$ where~$\lambda$ is an additional parameter and $n$ is the cardinality of the multispread.

Whereas spreads and multifold spreads are completely characterized up to parameters
(\cite{Hirschfeld79}, see Lemma~\ref{l:fold} below),
they are far from being characterized up to isomorphism, and only few research, e.g., see~\cite{MatTop:2009}, is done in this area due to obvious computational limits.
In the current paper, we consider the characterization problem
of admissible parameters of multispreads, which can be divided
into the following two questions.

\begin{problem}\label{prob:mu}
 Given $q$, $m$, and $t$, for which values of $\mu$ is there a $\param{\lambda}{\mu}{t}{m}{q}$-multispread, for some~$\lambda$?
\end{problem}
\begin{problem}\label{prob:lambda}
 For given $q$, $m$, $t$, and~$\mu$ such that there exists a~$\param{\lambda}{\mu}{t}{m}{q}$-multispread
 for some~$\lambda$, what is the minimum value of such~$\lambda$?
\end{problem}
As we will see in Section~\ref{s:1weight},
Problem~\ref{prob:mu} is essentially
the problem of characterization of admissible pairs (dimension, weight)
of $\FF_{q}$-linear (if $q$ is prime, additive)
one-weight codes over~$\FF_{q^t}$,
while Problem~\ref{prob:lambda} is essentially
the problem of characterization of the admissible triples (dimension, weight, length) of such codes.

\begin{remark}
    It should be noted that multispreads
    are equivalent to special multifold
    partitions of the space, in the sense 
    of~\cite{El-Zanati-et-al:lambda},
    where the dimension of the elements of a partition is bounded by~$t$ and 
    each subspace of dimension~$t-i$
    is included with multiplicity divisible by~$q^i$.
    However, motivated by the connection with codes,
    we are interested only in the existence of multispreads for given parameters $\para{\lambda}{\mu}$.
    For (multifold) space partitions, the main problem is to characterize all admissible types,
    where the type reflects the information about the number of subspaces of each dimension
    in a partition or a multifold partition.
    A~survey of results on partitions of a finite space can be found in~\cite{Heden:2012:survey}.
    Some of our results (Corollary~\ref{c:1fold}, \ref{Appendix:A2}) can also be considered as a contribution to that theory.
\end{remark}

The current study follows
a well-known approach relating optimal additive codes to geometric objects.
For example, linear codes that meet the Griesmer bound
and additive MDS codes are equivalently defined as minihypers
and arcs in projective spaces. 
In~\cite{BlokBrow:2004}, a geometric approach to finding the minimum distance of an additive code from its generator matrix is given.
Using these connections, 
the parameters of optimal additive $4$-ary codes of $\FF_2$-dimension up to~$6$
were characterized in~\cite{BMP:2021};
a class of optimal additive one-weight $4$-ary codes was constructed in~\cite{BMP:2024}. 
From the coding perspective, multispreads are
a geometric representation for additive one-weight codes and for additive completely regular codes with covering radius~$1$, see Section~\ref{s:1weight}.
Using this link, a complete characterization of the parameters of multifold $1$-perfect codes in $q$-ary Hamming graphs was obtained for all~$q$ being a power of a prime~\cite{Kro:mufold}.
The definitions and background on additive and surveys of completely regular codes are given in~\cite{Delsarte:1973}, \cite{KroPot:Ch1}, \cite{Zinoviev:Ch2}.

The paper is organized as follows.
In Section~\ref{s:1weight} we prove the connection of multispreads
with additive one-weight codes.
In Section~\ref{s:dual} we derive
the parameters of the multifold partition of the space 
dual to a multispread with given parameters.
We discuss three
special cases $\lambda=0$, $\mu<q$, and $t>m$ of multispreads in Section~\ref{s:special}.
Necessary conditions for the existence of multispreads
are considered in Section~\ref{s:ness}.
A constructive apparatus for multispreads is developed in Section~\ref{s:constr}.
Using the technique developed in the preceding sections,
in Section~\ref{s:param} we characterize, up to parameters,
some infinite families of multispreads, 
including all the cases when $t=2$,
$t=3$ and $q\in\{2,3\}$, $t=4$ and $q=2$. 
Equivalently,  the parameters of additive one-weight codes are characterized over fields of any prime-square order,
of order~$8$,  order~$27$,  order~$16$.

\section{%
Additive one-weight codes
over non-prime fields}\label{s:1weight}

\newcommand{\vc}[1]{\bar{#1}}
\newcommand{\wt}{\mathrm{wt}}

A Hamming space $H(n,Q)$ over the words of length~$n$ in an alphabet~$\Sigma$ of size~$Q$ is the metric space where the distance (Hamming distance) between two words is the number of positions in which they differ.
We consider the alphabet consisting of all tuples
of length~$t$ over~$\FF_\qp$,
where $\qp$ is a prime power, i.e., $\Sigma=\FF_\qp^t$,
and the Hamming space $H(n,\qp^t)$
has also the structure of the $nt$-dimensional
vector space $\Sigma^n=\FF_\qp^{nt}$ over~$\FF_\qp$.
An \emph{$\FF_\qp$-linear code}
(if $\qp$ is prime, an \emph{additive code};
if $t=1$, a \emph{linear code})
is a subspace of the vector space $\Sigma^n=\FF_\qp^{nt}$.
A \emph{generator matrix} of such a code is a matrix whose rows form a basis of the code.
A \emph{parity-check matrix} is a matrix whose rows form a basis of the subspace dual to the code.

\begin{remark}
    Alternatively, $\FF_\qp$-linear codes
    (additive if $\qp$ is prime)
    in $H(n,\qp^t)$
    can be treated as subsets of the vector space $\FF_{\qp^t}^n$ closed under addition and multiplication
    by a constant from the subfield~$\FF_{\qp}$ of~$\FF_{\qp^t}$. 
    There is no essential difference between the two representations of $\FF_\qp$-linear codes
    because $\FF_{\qp^t}$ is a $t$-dimensional
    vector space over~$\FF_{\qp}$, and the elements
    of~$\FF_{\qp^t}$ can be written as $t$-tuples of $\FF_{\qp}$-coordinates in some fixed basis. 
\end{remark}

The \emph{weight}, $\wt(\vc x)$ of a word $\vc x$ is the number of nonzero symbols in it.
Note that in our alphabet $\FF_\qp^t$ all vectors  except $(0,\ldots,0)$ are considered as nonzero symbols. 
An $\FF_\qp$-linear code 
of $\FF_\qp$-dimension~$m>0$ in $H(n,\qp^t)$, is called a \emph{one-weight $[n,m/t,d]_{\qp^t}$
code}
if all nonzero codewords have weight~$d$ in~$H(n,\qp^t)$.
Linear one-weight codes
are characterized
by Bonisoli's theorem~\cite{Bonisoli},
which can be treated in terms of multispreads as follows.
\begin{lemma}
\label{l:Bonisoli}
An $m\times n$ matrix over~$\FF_\qp$ with columns $h_1$, \ldots, $h_n$
is a generator matrix of a linear one-weight $[n,m,d]_\qp$ code
if and only
if $\{\langle h_1 \rangle, \ldots, \langle h_n \rangle\}$
is a $\param{\lambda}{\mu}{1}{m}{\qp}$-multispread, where 
\begin{equation}\label{eq:BonLM}
\mu = \frac{d}{\qp^{m-1}}, 
\qquad
\lambda = (\qp-1)n-\mu(\qp^m-1).
\end{equation}
\end{lemma}

We generalize this fact
to the wider class
of $\FF_\qp$-linear
one-weight codes 
over the alphabet~$\FF_\qp^t$.
\begin{theorem}\label{th:Bonisoli2}
An $m\times nt$ matrix~$M$ over~$\FF_\qp$ with columns
$h_{1,1}$, \ldots, $h_{1,t}$, $h_{2,1}$, \ldots, $h_{n,t}$
is a generator matrix of an $\FF_\qp$-linear
one-weight code in $H(n,\qp^t)$ if and only
if the collection $\{\langle h_{1,1}, \ldots, h_{1,t} \rangle,
\ldots,
\langle h_{n,1}, \ldots, h_{n,t} \rangle \} $
is a $\param{\lambda}{\mu}{t}{m}{\qp}$-multispread for some~$(\lambda,\mu)$.
Moreover,
the parameters of the multispread
and the parameters of the corresponding
$\FF_\qp$-linear one-weight $[n,m/t,w]_{\qp^t}$ code satisfy
the equations 
$$w=\mu \qp^{m-t},
\qquad
\lambda = (\qp^t-1)n-\mu(\qp^m-1).$$
\end{theorem}
\begin{proof}
We define a map~$\phi$ from
$\FF_\qp^t$ to $\FF_\qp^{\qp^t-1}$ as follows:
$$ \phi(x_1, \ldots , x_t) = (\alpha_1x_1+ \cdots + \alpha_t x_t)_{(\alpha_1,\ldots,\alpha_t)\in \FF_\qp^{t*}}.$$
We extend this map coordinate-wise to the
words from~$\Sigma^n$, $\Sigma=\FF_\qp^t$,
and row-wise to the matrix~$M$.
In particular, the $t$ columns
$h_{i,1}$, \ldots, $h_{i,t}$ of~$M$,
$i\in\{1,\ldots,n\}$,
correspond to the $\qp^t-1$ columns
$$h_{i,\vc{\alpha}} =\alpha_1 h_{i,1}+ \cdots + \alpha_t  h_{i,t},\qquad
\vc{\alpha}=(\alpha_1,\ldots,\alpha_t)\in \FF_\qp^{t*},$$
of~$\phi(M)$.
We deduce the following multiset identity:
\begin{equation}\label{eq:muid}
\qp^{t - \dim \langle h_{1,1}, \ldots, h_{1,t} \rangle } \times \langle h_{1,1}, \ldots, h_{1,t} \rangle - \{ 0 \} = 
\frac{1}{\qp-1}
\sum_{\vc{\alpha} \in \FF_\qp^{t*}}
 \big(\qp^{t - \dim \langle h_{i,\vc{\alpha}} \rangle } \times \langle h_{i,\vc{\alpha}} \rangle  - \{ 0 \} \big).
\end{equation}
Here, in each part of the equation,
the cardinality of the multiset is~$\qp^t-1$,
and each nonzero element of 
$\langle h_{1,1}, \ldots, h_{1,t} \rangle$ 
occurs with multiplicity $\qp^{t - \dim \langle h_{1,1}, \ldots, h_{1,t} \rangle }$.
From~\eqref{eq:muid} and the definition 
of a multispread, we get the following:
\begin{itemize}
\item[(i)]
\emph{%
$\{\langle h_{i,1}, \ldots, h_{i,t} \rangle: \ i\in\{1,\ldots,n\} \} $ 
is a $\para{\lambda}{\mu}$-multispread
if and only if \\
$\{\langle h_{i,\vc{\alpha}} \rangle: 
\ i\in\{1,\ldots,n\},\  {\vc{\alpha} \in \FF_\qp^{t*}}\} $ is a $\para{(\qp-1)\lambda}{(\qp-1)\mu}$-multispread.%
}
\end{itemize}

Next, for every
nonzero~$ x$ from~$\FF_\qp^t$, the value $\phi( x)$
has exactly $(\qp-1)\qp^{t-1}$ nonzero components.
Therefore, $\phi$ is a scaled isometry from
$H(n,\qp^t)$ to $H(n(\qp^t-1),\qp)$, that is,
$\wt(\phi(\vc x))=  (\qp-1)\qp^{t-1}\wt(\vc x)$ for
every~$\vc x$ from~$H(n,\qp^t)$. 
In particular, 
\begin{itemize}
\item[(ii)]
\emph{%
the matrix $M$ generates
an $\FF_\qp$-linear one-weight $[n,m/t,w]_{\qp^t}$ code
if and only if 
\linebreak[4]
$\phi(M)$ generates
a one-weight $[n(\qp^t-1),m,d]_{\qp}$ code, where $d=(\qp-1)\qp^{t-1} w$.
}
\end{itemize}
By Lemma~\ref{l:Bonisoli}, we can rewrite the second part of~(ii) as follows:
\begin{itemize}
\item[(iii)]
the matrix $M$ generates
an $\FF_\qp$-linear one-weight $[n,m/t,w]_{\qp^t}$ code
\emph{%
if and only if \linebreak[4]
$\{\langle h_{i,\vc{\alpha}}\rangle: 
\ i\in\{1,\ldots,n\},\  {\vc{\alpha} \in \FF_\qp^{t*}}\} $ is a $\para{{\lambda'}}{\mu'}$-multispread,
where 
$$\mu' = \frac{d}{\qp^{m-1}} 
= \frac{(\qp-1)w}{\qp^{m-t}} ,\qquad
\lambda' = (\qp-1)(\qp^t-1)n - \mu'(\qp^m-1).$$
}
\end{itemize}
Finally, by (i),
\begin{itemize}
\item[(iv)]
the matrix $M$ generates
an $\FF_\qp$-linear one-weight $[n,m/t,w]_{\qp^t}$ code
\emph{%
if and only if \linebreak[4]
$\{\langle h_{i,1}, \ldots, h_{i,t} \rangle: \ i\in\{1,\ldots,n\} \} $ 
is a $\para{\lambda}{\mu}$-multispread,
where
}
$$\mu = \frac{\mu'}{\qp-1}
      = \frac{w}{\qp^{m-t}},
\qquad
\lambda= \frac{\lambda'}{\qp-1} = (\qp^t-1)n - \mu(\qp^m-1).
$$
\end{itemize}
\end{proof}
\begin{example}
    Consider the matrix
\begin{equation*}\label{mat} M=\left(
\begin{array}{cc|cc|cc|cc|cc}
   1&0&1&0&1&0&0&0&1&0\\
   1&0&0&0&1&1&1&0&0&1\\
   1&0&0&1&0&1&0&1&0&0\\
\end{array}
\right),\end{equation*}
arising from the
$\param{1}{2}{2}{3}{2}$-multispread
in Fig.~\ref{f:fig1}.
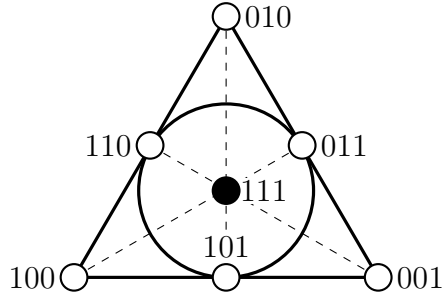
\begin{figure}[ht]
\centering
\begin{tikzpicture}[nd/.style={draw,circle,thick,fill=white,minimum size=10pt,inner sep=0pt,}]
\draw[very thick]  (0,0) -- (4,0);
\draw[very thick]  (0,0) -- (2,3.46);
\draw[very thick]  (2,3.46) -- (4,0);
\draw[very thick] (2,1.15) circle (1.15)  ;
 \draw [dashed] (0,0) -- (3,1.75);
\draw [dashed] (4,0) -- (1,1.75);
\draw [dashed] (2,0) -- (2,3.46); 
\draw [] (2,0.20) node [anchor=south,fill=white, inner sep=2pt, outer sep=0pt] {101};
\draw (2.15,1.15) node [anchor=west,fill=white, inner sep=1pt, outer sep=0pt] {111};
\node[nd,fill=black] at (2,1.15){};
\node[nd] at (3,1.75){};\draw (3.1,1.75) node [anchor=west] {011};
\node[nd] at (1,1.75){};\draw (0.9,1.75) node [anchor=east] {110};
\node[nd] at (2,3.46){};\draw (2.1,3.46) node [anchor=west] {010};
\node[nd] at (0,0){}; \draw (-0.1,0) node [anchor=east] {100};
\node[nd] at (2,0){};
\node[nd] at (4,0){}; \draw (4.1,0) node [anchor=west] {001};
\end{tikzpicture}
\caption{The  representation of a
$\param{1}{2}{2}{3}{2}$-multispread
via Fano plane. The multispread consists of the $1$-subspace $\langle 111\rangle$ (the black bullet) and four $2$-subspaces 
$\langle 100,001 \rangle$, 
$\langle 110,011 \rangle$, 
$\langle 010,001 \rangle$, 
$\langle 100,010 \rangle$ 
(solid lines).
}\label{f:fig1}
\end{figure}
The columns of the matrix are formed by
spanning sets of size $t=2$ for the $1$- and $2$-subspaces of the multispread. 
It is a generator matrix
of an additive one-weight $[5,1.5,4]_4$ code.
This code
 is viewed as a quaternary additive code of length~$5$ where any pair of bits in positions $i-1$, $i$ for  even~$i$ is treated as a single symbol from~$\FF_2^2$ in position $\frac{i}{2}$.
Alternatively, we can treat this code as a code of length~$5$ over $\FF_4$, if we map the pair of bits
$(a_{i-1},a_{i})$  to the element $a_{i-1}+xa_{i}$ of~$\FF_{2^2}=\FF_2[x]/\langle x^2+x+1\rangle$.
\end{example}
In view of Theorem~\ref{th:Bonisoli2},
the main results of the current paper can be treated as the characterization of
$\FF_\qp$-linear (additive, if $\qp$ is prime)
one-weight codes in the  cases listed in the following corollary. This
includes the complete characterization of parameters of additive one-weight codes over the alphabets of size $\qp^2$, $\qp^3\in\{8,27\}$, and~$\qp^4=16$.
\begin{corollary}
Assume a prime power~$\qp$
and positive integers $t$, $m$, $n$, and~$w$ satisfy one of the following conditions:
\begin{enumerate}
    \item[\rm(a)] $t=2$ {\rm(Theorem~\ref{th:t2});}
    \item[\rm(b)] $t=3$, $\qp\in \{2,3\}$ {\rm(Theorems~\ref{th:t3m4},~\ref{th:t=3m=4});}
    \item[\rm(c)] $t=4$, $\qp=2$, 
          $(m,w)\not\in \{(5,4),(5,6)\}$ {\rm(Theorems~\ref{th:t=4m=6},~\ref{th:t=4m=7},~\ref{th:t=4m=5});}
    \item[\rm(d)] $t=3$, $m\equiv 1 \bmod 3$ {{\rm(Theorem~\ref{th:t3m4});}}
    \item[\rm(e)] $t=4$, $m\equiv 2 \bmod 4$ {\rm(Theorem~\ref{th:t=4m=6});}
    \item[\rm(f)] $m\equiv 0 \bmod t$ {\rm(Proposition~\ref{p:m0});}
    \item[\rm(g)] $m<t$ {\rm(Proposition~\ref{p:t>m}).}
\end{enumerate}
Then an $\FF_\qp$-linear 
one-weight $[n,m/t,w]_{\qp^t}$ code
exists if and only if $$w=\qp^{m-t}\mu \quad \mbox{and} \quad
n\ge \frac{\qp^m-1}{\qp^t-1}\mu$$
for some positive integer~$\mu$ such that the following hold:
\begin{itemize}
    \item[\rm(i)] $\mu\ge \qp$ if $t$ does not divide~$m;$
    \item[\rm(ii)] $\qp^{t-m}$ divides $\mu$ if $m<t$.
\end{itemize}
Additive one-weight 
$[n,1.25,4]_{2^4}$ and $[n,1.25,6]_{2^4}$ codes do not exist for any~$n$ \rm(Corollary~\ref{c:l2452}).
\end{corollary}

A nonempty set $C$ of vertices of $H(n,Q)$ is called a \emph{completely regular code} of \emph{covering radius~$1$} with \emph{intersection array} $\{b;c\}$
if every vertex in~$C$ has exactly $b$ neighbors
not in~$C$ (and $n(Q-1)-b $ neighbors in~$C$)
and if every vertex not in~$C$
has exactly~$c$ neighbors in~$C$ (and $n(Q-1)-c$ neighbors not in~$C$). According to the Delsarte theory~\cite{Delsarte:1973}, an additive code is one-weight
if and only if its dual is a completely regular code of covering radius~$1$. From this point of view, Theorem~\ref{th:Bonisoli2} is the dual version of the
following theorem proved in~\cite{Kro:mufold}. However, giving the corresponding background to explain this relation would take much more space than the direct proof of Theorem~\ref{th:Bonisoli2}.
\begin{theorem}[{\cite[Theorem 1]{Kro:mufold}}]\label{t:original}
A $\param{\lambda}{\mu}{t}{m}{\qp}$-multispread~$S$ from $n$ subspaces of~$\FF_q^m$ exists if and only if in $H(n,\qp^t)$ 
there is an $\FF_\qp$-linear 
completely regular code with intersection array
$\{n(\qp^t-1)-\lambda;\mu\}$ and cardinality $\qp^{nt-m}$.
\end{theorem}
So, the results of the paper can also be treated as the characterization of classes of parameters of additive
completely regular codes of covering radius~$1$.
\section{Dual multifold partitions of a vector space}\label{s:dual}
By $\perp$, we denote an arbitrary fixed
orthogonality relation
on the vectors of~$\FF_q^m$.
A multiset $ S$ of subspaces of~$\FF_q^m$
is said to be a 
\emph{$\mu$-fold partition}
of the space~$\FF_q^m$ if every nonzero
vector of~$\FF_q^m$ lies in exactly~$\mu$ 
subspaces from~$ S$.
Straightforwardly from the definitions,
$\{C_1,\ldots,C_n\}$
is a $\param{\lambda}{\mu}{t}{m}{q}$-multispread
if and only if $\{\,\underbrace{C_1,...,C_1}_{\makebox[5mm]{$\scriptstyle q^{t-\dim(C_1)}\text{ times}$}},\,\ldots,\,\underbrace{C_n,...,C_n}_{\makebox[5mm]{$\scriptstyle q^{t-\dim(C_n)}\text{ times}$}}\,\}$
is a $\mu$-fold partition of the space.
In~\cite{El-Zanati-et-al:lambda},
it was shown that from a multifold partition
one can construct a dual multifold partition.
The following theorem is a special case of that duality.

\begin{theorem}\label{th:dual}
A multiset
$S=\{C_1,\ldots,C_n\}$
of subspaces of~$\FF_q^m$ having dimension at most~$t$
is a $\param{\lambda}{\mu}{t}{m}{q}$-multispread if and only if 
$\lambda + \mu(q^m-1) = n(q^t-1)$
and
$\{C_1^\perp,\ldots,C_n^\perp\}$
is a $\nu$-fold partition of~$\FF_q^m$, where
\begin{equation}\label{eq:dualN}
    \nu = n-q^{m-t}\mu 
\end{equation}
or, equivalently,
\begin{equation}\label{eq:dualL}
 (q^t-1) \nu = {(q^{m-t}-1) \mu+\lambda}.
\end{equation}
\end{theorem}
\begin{proof}
Comparing the cardinalities in the left and right parts of~\eqref{eq:def-mu}, we get
$$\lambda + \mu (q^m-1) = n (q^t-1)$$
for a $\param{\lambda}{\mu}{t}{m}{q}$-multispread of cardinality~$n$.
Therefore, \eqref{eq:dualN} and~\eqref{eq:dualL}
are equivalent.

Let $S=\{C_1,\ldots,C_n\}$
be a $\param{\lambda}{\mu}{t}{m}{q}$-multispread; denote $S^\perp =\{C_1^\perp,\ldots,C_n^\perp\}$. Consider a nonzero
vector~$x$ in~$\FF_q^m$ and count 
the number~$n_{x}$
of subspaces in $S^\perp $
that \emph{do not contain}~$x$. 
Equivalently, $n_{x}$ is the number of subspaces in $S $
that are not orthogonal to~$x$.
Each such subspace~$U$ contains
exactly $q^{\dim(U)-1}$ vectors
orthogonal to~$x$ and 
\linebreak[4]
$q^{\dim(U)}-q^{\dim(U)-1}$ vectors
not orthogonal to~$x$.
It follows that the multiset
$q^{t-\dim(U)}\times U$
contains $q^t-q^{t-1}$ such vectors
and the multiset in the right part
of~\eqref{eq:def-mu} contains
exactly $n_{x} (q^t-q^{t-1})$ vectors
not orthogonal to~$x$.
On the other hand, the left part of~\eqref{eq:def-mu} contains
exactly $\mu(q^m-q^{m-1})$ such vectors.
We conclude that $n_{x}=\frac{\mu(q^m-q^{m-1})}{q^t-q^{t-1}}= \mu q^{m-t}$.
Therefore, the number
of subspaces in $S^\perp $
that \emph{contain}~$x$ is 
$n-n_{x}=n-\mu q^{m-t}$, 
which is exactly 
the~$\nu$ 
from~\eqref{eq:dualN}. By the definition, $S^\perp $ is a $\nu$-fold partition of~$\FF_q^m$.

Now, let $S^\perp = \{C^\perp_1,\ldots,C^\perp_n\}$
be a $\nu$-fold partition of~$\FF_q^m$ into subspaces of dimension at least $m-t$. 
In particular, we have $$\sum_{V\in S^\perp}(q^{\dim(V)}-1)=\nu(q^m-1)$$
and 
\begin{equation}\label{eq:V}
   \sum\limits_{V\in S^\perp}q^{\dim(V)} =  \nu (q^m-1) + n.
\end{equation}
For a nonzero vector~$x$,
a subspace~$V$ that is \emph{not orthogonal} to~$x$
contains exactly \linebreak[4]\mbox{$q^{\dim(V)}(1-q^{-1})$} vectors not orthogonal to~$x$.
Since there are \mbox{$q^m(1-q^{-1})$} such vectors in~$\FF_q^m$, from the 
$\nu$-fold partition definition
we have
\begin{equation}\label{eq:Vp}
   \sum\limits_{V\in S^\perp:\ x\not\perp V}q^{\dim(V)} =  \nu q^m.
\end{equation}
Now, we
expand the right part in~\eqref{eq:def-lm} and confirm that $S$
satisfies the definition of a multispread:
\begin{multline*}\label{eq:igy}
\sum\limits_{U\in S:\ x\in U}q^{t-\dim(U)}
\stackrel{(V=U^\perp)} =
\sum\limits_{V\in S^\perp:\ x\perp V}q^{t-m+\dim(V)}
\\
=
\sum\limits_{V\in S^\perp}q^{t-m+\dim(V)}
-\sum\limits_{V\in S^\perp:\ x\not\perp V}q^{t-m+\dim(V)} 
\\
\stackrel{\eqref{eq:V},\eqref{eq:Vp}}=
q^{t-m} 
\big( \nu (q^m-1) + n - \nu q^m \big) = q^{t-m} (n-\nu)
\stackrel{\eqref{eq:dualN}}=
\mu.
\end{multline*}
\end{proof}
\section{Special cases}\label{s:special}

Here, we consider three special cases of multispreads,
for which the characterization
up to parameters follows
from simple arguments or known results. The first case,
$\lambda=0$,
corresponds to the known concept of multifold spread.
The second case, $\mu<q$, is reduced to the first one.
The third case is $t>m$.

A {\it $\mu$-fold spread} is a collection of $t$-subspaces such that every nonzero vector
of the space is in exactly~$\mu$ subspaces from the collection; 
so, by the definition, it is just a
$\param{0}{\mu}{t}{m}{}$-multispread.
The characterization of parameters of $\mu$-fold spreads is
 more or less simple;
 for completeness, we include the proof.

\begin{lemma}[on $\mu$-fold spreads {\cite[p.83]{Hirschfeld79}}, {\cite[Corollary~8]{El-Zanati-et-al:lambda}}]
\label{l:fold}
    A $\mu$-fold spread
    of $t$-subspaces
    of $\FF_q^m$ exists if and only if
     $t\le m$ and
    $\mu$ is divisible by
    $\frac{q^t-1}{q^s-1}$, where
    $s = \gcd(t,m)$.
\end{lemma}
\begin{proof}
\emph{If.} It is sufficient
to construct
a $\mu$-fold spread for $\mu = \frac{q^t-1}{q^s-1}$.
Consider the following notations: $s = \gcd(t,m)$,
$T=t/s$, $M=m/s$,
$Q=q^s$.
Let $C$ be a
 $T$-subspace  of~$\FF_Q^M$
(it exists because $t\le m$ and hence $T\le M$). Consider the structure of the finite field $\FF_{Q^M}$  on $\FF_Q^M$ and a complete system
$\bar\alpha =\{\alpha_1,\ldots,\alpha_{\frac{Q^M-1}{Q-1}}\}$ of pairwise linearly independent  representatives of~$\FF_Q^M$.
Here,
$S =\{\alpha_1 C,\ldots,\alpha_{\frac{Q^M-1}{Q-1}}C\}$ is a $\frac{Q^T-1}{Q-1}$-fold
spread of $\FF_Q^M$. It remains to note
that $\frac{Q^T-1}{Q-1}=\frac{q^t-1}{q^s-1}$, $\FF_Q^M$ is an $m$-dimensional
$\FF_q$-space, and $C$ (as well as all elements of $S$)
is its $t$-subspace over~$\FF_q$.

\emph{Only if.}
It is necessary that 
$t\le m$. Here,
$\mu(q^m-1)$ is divisible by $(q^t-1)$,
i.e., $\mu$ is divisible by
$\frac{q^t-1}{\gcd(q^t-1,q^m-1)}$.
It remains to show that
\begin{equation}\label{eq:gcd}
   \gcd(q^t-1,q^m-1) = q^{\gcd(t,m)}-1.
\end{equation}
We will prove it by induction on~$\max(t,m)$.
If $t=m=\gcd(t,m)$,
then \eqref{eq:gcd} is trivial.
Otherwise, one of $t$, $m$
is larger than the other, say $m>t$.
In this case,
$$q^m-1=q^{m-t}(q^t-1)+q^{m-t}-1\equiv q^{m-t}-1
\bmod q^{t}-1,$$
and we have
$$\gcd(q^t-1,q^m-1)=\gcd(q^t-1,q^{m-t}-1).$$
By the induction hypothesis,
the right part equals~$q^{\gcd(t,m-t)}-1$,
which is $q^{\gcd(t,m)}-1$.
\end{proof}

\begin{proposition}\label{p:mu<q}
    A $\param{\lambda}{\mu}{t}{m}{q}$-multispread
    such that $1\le\mu<q$
    exists if and only if
    $$ 
    t\le m, 
    \qquad
    \frac{q^t-1}{q^{\gcd(t,m)}-1} \,\Big|\, \mu,
    \qquad\mbox{and}\quad
    q^t-1 \,\big|\, \lambda.
    $$
\end{proposition}
\begin{proof}
 If $\mu<q$,
 then a $\param{\lambda}{\mu}{t}{m}{}$-multispread~$S$
  consists of only
 $0$- and $t$-subspaces.
 Indeed, a subspace of any other dimension cannot belong to~$S$, because it
 contains a nonzero vector of multiplicity at least~$q$, which is larger than~$\mu$. 
Hence, the collection of
$t$-subspaces from~$S$ forms
a $\para{0}{\mu}$-multispread
(discussed in Lemma~\ref{l:fold}),
while the $0$-subspaces from~$S$  form
a~$\param{\lambda=l(q^t-1)}{0}{t}{m}{}$-multispread,
where $l$ is the number of such subspaces.
\end{proof}

\begin{proposition}\label{p:t>m}
 A $\param{\lambda}{\mu}{t}{m}{q}$-multispread, $t>m$,
exists if and only if
\begin{equation}\label{eq:t>m:l}
q^{t-m} \, \big| \, \mu
\qquad\mbox{and}\quad
  \lambda = (q^{t-m}-1)\frac{\mu}{q^{t-m}}+l(q^t-1)
\end{equation}
for some nonnegative integer~$l$.
\end{proposition}
\begin{proof}

\emph{Necessity.} Since $t>m$, the dimension of any subspace
of~$\FF_q^m$
is not greater than~$m$ and each term in $\mu=\sum\limits_{U\in S,\mbox{ }x\in U}q^{t-\dim(U)}$ is divisible by $q^{t-m}$, hence $\mu$ is divisible by~$q^{t-m}$.
Moreover, we have at least $\frac{\mu}{q^{t-m}}$
subspaces in a $\param{\lambda}{\mu}{t}{m}{q}$-multispread and therefore
\linebreak[4]
$\lambda=\sum\limits_{U\in S}(q^{t-\dim(U)}-1)$ is at least
\mbox{$(q^{t-m}-1)\frac{\mu}{q^{t-m}}$}.
Therefore, $l\ge 0$ in
expression~\eqref{eq:t>m:l} for~$\lambda$.
Finally, rewriting~\eqref{eq:t>m:l} as
$$
\lambda
=-\mu(q^m-1) + \frac {\mu}{q^{t-m}}(q^t-1)+l(q^t-1),
$$
we see that $l$ is an integer
by Proposition~\ref{c:ness}.

\emph{Sufficiency.} Consider the multispread that consists of
the space $\FF_q^m$ with multiplicity $\mu/q^{t-m}$
and the trivial subspace of dimension~$0$ with multiplicity~$l$.
\end{proof}

\section{Necessary conditions}\label{s:ness}

Assume that there exists
a $\param{\lambda}{\mu}{t}{m}{q}$-multispread~$S$
from $n$ subspaces of~$\FF_q^m$,
where $t\le m$ and $\mu\ge q$.
In general, it can consist of subspaces of different
dimensions from~$0$ to~$t$,
and depending on the dimension,
each such subspace makes different contributions
to the $\lambda$- and $\mu$-parts of the multiset
equation~\eqref{eq:def-mu}.
However,
in total, each subspace~$U$
brings $|q^{t-\dim(U)}\times U -\{0\}| = q^t-1$
vectors to the right part, and equating
the cardinalities of the left and right parts,
we get
$$ \lambda + \mu (q^m-1) = n (q^t-1), $$
which gives the following necessary condition
for $\lambda$, given $q$, $m$, $t$, and~$\mu$.
\begin{proposition}
\label{c:ness}
 If a $\param{\lambda}{\mu}{t}{m}{q}$-multispread exists, then
\begin{equation}\label{eq:mod}
    \lambda \equiv -\mu(q^m-1) \bmod q^t-1.
    \end{equation}
\end{proposition}

Next, we are now going to compare the $\mu$-component of
equation~\eqref{eq:def-mu},
i.e., the number of nonzero vectors
in the left and right parts of the equation.
Each subspace of dimension~$t-i$ brings
$q^t-q^i$ nonzero vectors
to the right part of~\eqref{eq:def-mu}. So, we get
\begin{equation}\label{eq:lr-mu}
 \mu(q^m-1)
 = \sum_{U\in S} (q^t-q^{t-\dim(U)})
 = \sum_{i=0}^t b_i (q^t-q^i),
\end{equation}
where $b_i$ is the number of $(t-i)$-subspaces
 in~$S$, $i=0,\ldots,t$.
Now we make two important observations.
At first, $0$-subspaces do not contribute to
\eqref{eq:lr-mu} because $q^t-q^i=0$ if $i=t$.
At second, if $S$ contains a subspace of dimension~$t-i>0$,
then $q^i\le \mu$, because such a subspace
has a nonzero vector, which comes to the right part of~\eqref{eq:def-mu}
with multiplicity at least~$q^i$,
while in the left part it has multiplicity~$\mu$.
Hence, $b_i = 0$ if $\log_q(\mu)<i<t$.
We immediately get the following necessary condition for $\mu$.

\begin{proposition}
\label{p:ness}
If a $\param{\lambda}{\mu}{t}{m}{q}$-multispread exists,
then there are positive integers
$b_0$, \ldots, $b_{i_{\mathrm{mx}}}$,
where ${i_{\mathrm{mx}}}={\min\{\lfloor\log_q(\mu)\rfloor,t-1\}} $, such that
\begin{equation}\label{eq:ness-mu}
 \mu(q^m-1)
 = \sum_{i=0}^{i_{\mathrm{mx}}} b_i (q^t-q^i).
\end{equation}
\end{proposition}
This condition always holds if the left part of~\eqref{eq:ness-mu}  is sufficiently large (for given~$q$ and~$t$).
However, as we see below,
for small parameters it is essential.
Before we show such examples,
let us deduce an easier-to-check corollary.
\begin{corollary}\label{p:ness-}
If a $\param{\lambda}{\mu}{t}{m}{q}$-multispread exists,
then there is an integer $n_0$ such that
$$
\mu \frac{q^m-1}{q^t-1}\le n_0 \le \mu \frac{q^m-1}{q^t-q^{i_{\mathrm{mx}}}},
\qquad\mbox{where ${i_{\mathrm{mx}}}={\min\{\lfloor\log_q(\mu)\rfloor,t-1\}} $.}
$$
\end{corollary}
\begin{proof}
 Taking $\displaystyle n_0 = \sum_{i=0}^{i_{\mathrm{mx}}} b_i$,
 we see that both inequalities
 $ \mu(q^m-1) \le n_0 (q^t-1) $ and
 \linebreak[4]
 $ \mu(q^m-1) \ge n_0 (q^t-q^{i_{\mathrm{mx}}}) $
 follow from~\eqref{eq:ness-mu}.
\end{proof}

\begin{corollary}\label{c:l2452}
 For any $\lambda$,
 there are no
 $\param{\lambda}{2}{4}{5}{2}$- and
 $\param{\lambda}{3}{4}{5}{2}$-multispreads.
\end{corollary}
\begin{proof}
 In both cases, ${i_{\mathrm{mx}}}=1$, and by Corollary~\ref{p:ness-} we get
 $$4< 2\cdot\frac{31}{15} \le n_0 \le 2\cdot\frac{31}{14}<5
 \quad\mbox{and}\quad
 6< 3\cdot\frac{31}{15} \le n_0 \le 3\cdot\frac{31}{14}<7,$$
 respectively.
 In each case, there is no such integer~$n_0$.
\end{proof}

\section{Constructions}\label{s:constr}
\subsection{Basic  constructions}\label{s:basic}

\begin{lemma}\label{l:sum}
    The union of a $\param{\lambda}{\mu}{t}{m}{q}$-multispread and
a    $\param{\lambda'}{\mu'}{t}{m}{q}$-multispread is a \linebreak[5]
$\param{\lambda+\lambda'}{\mu+\mu'}{t}{m}{q}$-multispread.
\end{lemma}

\begin{lemma}\label{l:rec}
Let $S$ be a $\param{\lambda}{\mu}{t}{m}{q}$-multispread $S$.
Then
there is a $\param{\lambda'}{\mu'}{t'}{m'}{q}$-multispread
where
\begin{itemize}
\item[\rm (a)]
    $m'=m$,\quad
    $t'=t$,\quad
    $\lambda'=\lambda+q^t-1$,\quad
    $\mu'=\mu;$
\item[\rm (b)]
    $m'=m$,\quad
    $t'=t$,\quad
    $\lambda'=\lambda$,\quad
    $\mu'=\mu+\frac{q^t-1}{q^s-1}$,
    where $s=\gcd(t,m);$
\item[\rm (c)]
    $m'=m+t$,\quad
    $t'=t$,\quad
    $\lambda'=\lambda$,\quad
    $\mu'=\mu;$
\item[\rm (d)]
    $m'=m-1$,\quad
    $t'=t$,\quad
    $\lambda'=\lambda+(q-1)\mu$,\quad
    $\mu'=q\mu;$
\item[\rm (e)]
    $m'=m$,\quad
    $t'=t+1$,\quad
    $\lambda'=q\lambda+(q-1)|S|$,\quad
    $\mu'=q\mu.$
\end{itemize}
Moreover, if $q=p^l$ for some prime power $p$ and integer~$l$,
then
\begin{itemize}
\item[\rm (f)]
there is a $\param{\lambda}{\mu}{lt}{lm}{p}$-multispread.
\end{itemize}
\end{lemma}
\begin{proof}
(a) This follows from Lemma~\ref{l:sum} and
the existence of a trivial $\param{q^t-1}{0}{t}{m}{q}$-multispread,
consisting of one $0$-subspace.

(b) This follows from Lemmas~\ref{l:fold} and~\ref{l:sum}.

(c) We consider two cases, $t\le m$ and $t > m$.

1. Let $t\le m$.
We first  describe the partition
from~\cite{Bu:1980:pertitions}
of an $(m+t)$-space over~$\FF_q$
into one $m$-subspace
     and $q^m$ subspaces of dimension~$t$.

     The Galois field $\FF_{q^m}$
     is itself an $m$-dimensional
     vector space over~$\FF_q$.
     Denote by~$U$ one of its $t$-subspaces.
     The $(m+t)$-dimensional vector space
     (over~$\FF_{q}$)
     \begin{equation}
         \label{eq:mtspace}
\{ (\beta| u) :\ \beta\in\FF_{q^m} ,\,u \in U \},
     \end{equation}
is partitioned into the $m$-subspace
     \begin{equation}
         \label{eq:msspace}
\{(\beta|{\boldsymbol 0}):\ \beta \in \FF_{q^m}\}
     \end{equation}
and the $t$-subspaces
\begin{equation}
         \label{eq:tsspaces}
         U_{\alpha} = \{ (\alpha u| u) :\ u \in U \}, \qquad \alpha \in\FF_{q^m}.
\end{equation}
     Since all $(m+t)$-dimensional
     vector spaces over the same field are isomorphic,
     we also have a required vector-space partition
     of~$\FF_{q}^{m+t}$.

Now, having a
$\param{\lambda}{\mu}{t}{m}{}$-multispread
in~\eqref{eq:msspace},
we can easily complete it to a
$\param{\lambda}{\mu}{t}{m+t}{}$-multispread
in~\eqref{eq:mtspace} by adding
each subspace from~\eqref{eq:tsspaces}
$\mu$ times.
Since all vector spaces of the same
dimension over the same field are isomorphic,
we can also construct a
$\param{\lambda}{\mu}{t}{m+t}{q}$-multispread
in $\FF_q^{m+t}$.

2. Let $t > m$.
By Proposition~\ref{p:t>m}, $\mu$ is a multiple of~$q^{t-m}$ and $\lambda$ 
has form~\eqref{eq:t>m:l}, for some~$l$.
Then a required 
$\param{\lambda}{\mu}{t}{m+t}{q}$-multispread
can be obtained
by constructing
a 
$\param{l(q^t-1)}{\frac{\mu}{q^{t-m}}}{t}{2t}{q}$-multispread as the union of $l$ 
$\param{q^t-1}{0}{t}{2t}{q}$-multispreads (see p.(a) of this lemma) and 
$\frac{\mu}{q^{t-m}}$
$\param{0}{1}{t}{2t}{q}$-multispreads (see p.(b)) and then projecting $t-m$ times (see p.(d)).

(d) By removing the last coordinate position,
we project every vector of~$\FF_q^m$ to a vector of~$\FF_q^{m-1}$.
Under this operation, every subspace
 of~$\FF_q^m$ is mapped to
a subspace of~$\FF_q^{m-1}$ of the same or smaller
dimension.
 Since the preimage of every nonzero vector
 in~$\FF_q^{m-1}$ is $q$ nonzero vectors in~$\FF_q^{m}$, we get $\mu'=q\mu$.
 Since the preimage of the zero vector
 in~$\FF_q^{m-1}$ is the zero vector and $q-1$ nonzero vectors in~$\FF_q^{m}$, we get
$\lambda'=\lambda+(q-1)\mu$.

(e) We treat
    each subspace of pseudodimension $t$
     as a subspace with pseudodimension $t+1$.
     The corresponding values for $\lambda'$ and $\mu'$ follow directly from the definition.

(f) Similarly to (e), we see that the original
$\param{\lambda}{\mu}{t}{m}{q}$-multispread
is itself a multispread with required parameters,
since the $m$-dimensional space and $t$-subspaces over~$\FF_q$
are respectively
an $lm$-dimensional space and $lt$-subspaces over~$\FF_p$ if $q=p^l$.
\end{proof}
\begin{corollary}\label{c:suff}
Given $q$, $t$, $m_0$, $\mu_0\ge q$,
assume that a $\param{\lambda_{\min}}{\mu}{t}{m}{q}$-multispread exists
for $m=m_0$ and all~$\mu$ from the following interval:
 $$\mu_0 \leq \mu < \mu_0 + \frac{q^t-1}{q^{s}-1}, \qquad s=\gcd(t,m) ,$$
 where $\lambda_{\min} = \lambda_{\min}(q,m,t,\mu)$ is the minimum~$\lambda$
 satisfying~\eqref{eq:mod}.
 Then a $\param{\lambda}{\mu}{t}{m}{q}$-multispread exists
 for all $m\geq m_0$, $m \equiv m_0\bmod t$, 
  all $\mu \ge \mu_0$, and all~$\lambda$ satisfying~\eqref{eq:mod}.
 \end{corollary}

\begin{proof}
Since $q^s-1$ divides $q^m-1$, we have
$$
\Big(\mu+\frac{q^t-1}{q^s-1}\Big)(q^m-1)
\equiv
\mu(q^m-1)
\bmod q^t-1
$$
and see from~\eqref{eq:mod} that $\lambda_{\min}$ does not change
if we increase~$\mu$ by the increment of~$\frac{q^t-1}{q^s-1}$.
So, by Lemma~\ref{l:rec}(b), we have the existence of a $\param{\lambda_{\min}}{\mu}{t}{m}{q}$-multispread for $m=m_0$ and all~$\mu\ge \mu_0$.

Next, we observe that
$$
\mu(q^{m+t}-1) = \mu q^{m} (q^t-1) + \mu(q^m-1)
\equiv
\mu(q^m-1)
\bmod q^t-1
$$
and hence increasing~$m$ by~$t$ does not change~$\lambda_{\min}$ as well.
By Lemma~\ref{l:rec}(c), we have the existence of a $\param{\lambda_{\min}}{\mu}{t}{m}{q}$-multispread for all $m\geq m_0$, $m \equiv m_0\bmod t$, and all~$\mu\ge \mu_0$.

Finally, with Lemma~\ref{l:rec}(a), we expand the existence to all $\lambda$ satisfying~\eqref{eq:mod}.
\end{proof}

With Corollary~\ref{c:suff}, constructing a finite number of multispreads
can be sufficient to characterize all admissible multispread parameters
with given~$q$ and~$t$ (in particular, this would yield the characterization
of all parameters of additive one-weight $q^t$-ary codes).
The best case is if the hypothesis of Corollary~\ref{c:suff} holds
with $\mu_0=q$ and $m_0 \le 2t$. We will see that it is true, for example,
if $t=2$. However, it cannot be true always, as shown in Corollary~\ref{c:l2452}.

\subsection{Switching constructions}\label{s:switch}

The next two lemmas show that sometimes we can change the parameter~$\mu$
of a multispread by replacing some special subset in it with some other subset. Such operation is often called a switching, but usually
by a switching one means a replacement that does not change the main parameters of considered configurations, while in our case $\mu$ is changed.

\begin{lemma}\label{l:t+1}
    Assume that $m=t+s$
    and there exists
    a $\param{\lambda}{\mu}{t}{m}{q}$-multispread
    $\bar S$
    that contains a $(t-s)$-subspace.
    Then there exists a
    $\param{\lambda-(q^s-1)}{\mu+1}{t}{m}{q}$-multispread. 
\end{lemma}

\begin{proof}
Let $T$ be a $(t-s)$-subspace in  $\bar S$. We seek for $t$-subspaces $L_1,\ldots,L_{q^s+1}$ of $\FF_q^m$  such that their pairwise intersection is $T$ and the complement of $T$ to $\FF_q^{t+s}$ is covered by these subspaces with multiplicity~$1$.

Such a collection of $t$-subspaces always exists. Indeed, $\FF_q^{t+s}$
could be viewed as $T\oplus\FF_q^{2s}$.
The space $\FF_q^{2s}$ contains a spread of $s$-subspaces, see Lemma~\ref{l:fold}. The direct sums of these subspaces and the subspace~$T$ form a desired collection $L_1,\ldots, L_{q^s+1}$.

We now consider the multiset $S={\bar S}\setminus T \cup \bigcup\limits_{i=1,\ldots,q^s+1} L_i$.
The subspaces $L_1,\ldots, L_{q^s+1}$ cover each nonzero vector in $T$ with multiplicity $q^s+1$ and the remaining nonzero vectors with multiplicity~$1$.  In the multispread~$\bar{S}$,  the  subspace~$T$ accounts for $q^s$ in~$\mu$ only for nonzero vectors of~$T$. So after switching,  we see that $\mu$ is a constant increased by $1$ in $S$ and $S$ is a multispread. Finally, we note that
the $(t-s)$-subspace $T$ accounts for $q^s-1$ in $\lambda$ in ${\bar S}$, so the parameter $\lambda$ in~$S$ is reduced by $q^s-1$ compared to that of $\overline{S}$.
\end{proof}

The operation described in the proof of Lemma~\ref{l:t+1} can be reversed,
and we obtain the following.

\begin{lemma}\label{l:t-1}
    Let $\overline{S}$  be
    a $\param{\lambda}{\mu}{t}{m}{q}$-multispread, where $m=t+s$.
    If there is a $(t-s)$-subspace~$T$ such that $\bar S$ contains
    $q^s+1$ distinct $t$-subspaces whose pairwise intersection
    is~$T$,
    then replacing all these $q^s+1$ subspaces by~$T$ in~$\overline{S}$
    results in a
    $\param{\lambda+(q^s-1)}{\mu-1}{t}{m}{q}$-multispread.
\end{lemma}



\subsection{Multispreads from the Desarguesian spread
in \texorpdfstring{$\FF_q^6$}{Fq6}} \label{s:desarg}
The Desarguesian
spread of $3$-subspaces in $\FF_q^6$, see Lemma \ref{l:fold},  consists of the $q^3+1$ multiplicative cosets of $\FF_{q^3}$ in $\FF_{q^6}$:
$$a_0\FF_{q^3},  a_1 \FF_{q^3}, \ldots,a_{q^3}\FF_{q^3}.$$

Consider all possible intersections of a $4$-subspace $T$ of $\FF_q^6$ with
 the $3$-subspaces from the spread. Since a $4$-subspace and a $3$-subspace of $\FF_q^6$  necessarily have a  common nonzero vector, we have only the following two cases:
\begin{itemize}
    \item [(A)] the $4$-subspace $T$ includes one of $3$-subspaces from the spread and meets each of the remaining $q^3$ spread subspaces in a $1$-subspace;
    \item[(B)] the $4$-subspace $T$ meets $q+1$ of
    spread $3$-subspaces in $2$-subspaces and meets each of the remaining $q^3-q$ spread $3$-subspaces in a $1$-subspace.
\end{itemize}
Let $T$ be a $4$-subspace that satisfies~(B). We define {\it the block} of~$T$ to be the set of all $3$-subspaces from the Desarguesian spread that meet~$T$ in $2$-subspaces.

\def\Sc#1{\mathrm{Sc}_{#1}}

Throughout this section, by $\alpha$ we denote a primitive element of $\FF_{q^6}$. Consider the Singer cycle that acts on the elements of $\FF_{q^6}$ as follows:
$$x\rightarrow \alpha x ,$$
$x\in \FF_{q^6}$, and denote the cyclic group generated by this mapping as $\Sc{q^6}$.
The element $\alpha^{q^3+1}$ is a primitive element of
the subfield $\FF_{q^3}$, and we denote the subgroup of $\Sc{q^6}$ generated by $x\rightarrow \alpha^{q^3+1} x$ as $\Sc{q^3}$.
For a subspace~$T$ of~$\FF_q^6$, let $O_{q^3}(T)$ and $O_{q^6}(T)$ be the  orbits of $T$ under the action of $\Sc{q^3}$ and $\Sc{q^6}$, respectively.
\newcommand\Tr[1]{\mathrm{Tr}_{q^{#1}}}Denote by $\Tr{2}$ the trace map 
from $\FF_{q^6}$ to $\FF_{q^2}$, i.e., $\Tr{2}(x)=x + x^{q^2} + x^{q^4}$.

\begin{lemma}\label{l:action}
\begin{itemize}
 \item[\rm(i)]
For any $i\in\{0,\ldots,q^3\}$, the group $\Sc{q^3}$ fixes
 the multiplicative coset $a_i\FF_{q^3}$ in $\FF_{q^6}$ and the actions of $\Sc{q^3}$ on the $1$-subspaces and the $2$-subspaces of $a_i\FF_{q^3}$ are transitive.

\item[\rm(ii)]
Let $T$ be a $4$-subspace  satisfying condition (B).
Then all $\frac{q^3-1}{q-1}$ subspaces from $O_{q^3}(T)$  fulfill condition (B) with the same block.
Each nonzero vector of each
subspace from the block of~$T$
is in exactly $q+1$  subspaces from $O_{q^3}(T);$  each other nonzero vector of $\FF_{q^6}$ is in exactly one subspace from $O_{q^3}(T)$.

\item[\rm(iii)]
Let $T$ be $\{x \in \FF_{q^6}:\ \Tr{2}(x)=0 \}$. 
Then $T$ fulfills condition (B) and in the orbit $O_{q^6}(T)$ there are
$q^2-q+1 $
subspaces $T_1,\ldots, T_{q^2-q+1}$ with pairwise disjoint blocks.
\end{itemize}
\end{lemma}
\begin{proof}
(i) is straightforward.



(ii)   Let $T$
 fulfill Condition~(B).
From (i) we see that
the orbit $O_{q^3}(T)$ of $T$ under the action of $\Sc{q^3}$ consists of exactly $\frac{q^3-1}{q-1}$ subspaces fulfilling Condition (B) and moreover, the blocks of the subspaces in the orbit coincide.
Because each nonzero vector of $\FF_q^3$
is in $q+1$ $2$-subspaces of $\FF_q^3$,
each nonzero vector from any block subspace
is in exactly $q+1$ subspaces from $O_{q^3}(T)$.
Similarly, each nonzero vector of $\FF_q^3$ is in exactly one $1$-subspace, so each  nonzero vector from a non-block subspace is covered exactly once by the subspaces from~$O_{q^3}(T)$.

(iii) Let $T$ be $\{x \in \FF_{q^6}:\ \Tr{2}(x)=0 \}$. Note that $\FF_{q^6}$ is a $2$-dimensional
space over~$\FF_{q^3}$ and 
the Desarguesian spread consists of 
its $1$-dimensional $\FF_{q^3}$-subspaces.  
It is well known that $\Tr{k}$ is a $\FF_{q^k}$-linear mapping; therefore, $T$ is a $\FF_{q^2}$-subspace of $\FF_{q^6}$ and cannot include a nontrivial $\FF_{q^3}$-subspace (because the $\FF_{q^2}$-closure of a $\FF_{q^3}$-subspace is~$\FF_{q^6}$). We conclude that $T$
 fulfills~(B).
 As~$\beta T=T$
 for any nonzero $\beta$ in~$\FF_{q^2}$,
there are not more than 
$$\frac{q^{6}-1}{q^2-1}=(q^2+q+1)(q^2-q+1)$$ different subspaces in $O_{q^6}(T).$

On the other hand, by (ii), all $\frac{q^{3}-1}{q-1}=q^2+q+1$
subspaces from~$O_{q^3}(T)$ have the same block,
and hence the subspaces in $O_{q^6}(T)$
have not more than $q^2-q+1$ different blocks.

It remains to observe that these $q^2-q+1$ (or less) blocks, each of size $q+1$,
must cover all the $q^3+1$ subspaces in the Desarguesian spread,
because $\Sc{q^6}$ acts transitively on
the nonzero vectors of~$\FF_q^6$.
Since $q^3+1=(q+1)(q^2-q+1)$,
this covering is a partition.
\end{proof}

\begin{theorem}\label{th:des}Let
 $D=\{a_i\FF_{q^3}:\ i\in\{0,\ldots,q^3\}\}$ be the Desarguesian spread of $3$-subspaces in $\FF_q^6$ and
$T_1$, \ldots, $T_s$, $s\leq q^2-q+1$, be   $4$-subspaces from Lemma~\ref{l:action}(iii).
Let
$$
D'=\{U\in D:\ \dim(U\cap T_j)=1 \mbox{ for all }j\in\{1,\ldots,s\}\}.$$
 Then $D' \cup \bigcup\limits_{j=1,\ldots,s} O(T_j)$ is a
 $\param{\lambda}{q+s}{4}{6}{}$-multispread,
 $\lambda=(q^2-q+1-s)(q^2-1)$.
\end{theorem}
\begin{proof}

By Lemma \ref{l:action}(iii), there are $q^2-q+1$ subspaces $T_1,\ldots, T_{q^2-q+1}$ fulfilling $(B)$ with pairwise disjoint blocks.
From  Lemma \ref{l:action}(ii) we see that
any nonzero vector from any subspace of $D'$ and any  vector not belonging to any subspace from $D'$  respectively
    are exactly in $s$ and in $q+1+s-1=q+s$ subspaces from  $\bigcup\limits_{j=1,\ldots,s} O(T_j)$ respectively.
The collection of subspaces $D'$   ``levels up'' the multiplicity~$\mu$ up to $q+s$ for nonzero vectors from subspaces of~$D'$. Note that $\lambda=(q-1)(q^3+1-s(q+1))=(q^2-q+1-s)(q^2-1)$ because we have exactly $q^3+1-s(q+1)$ subspaces of dimension $3=t-1$  in~$D'$.
\end{proof}

For the dual multifold partitions, 
by Theorem~\ref{th:dual}, we obtain the following: 
\begin{corollary}\label{c:1fold}
For every $s\le q^2-q+1$,
there exists a ($1$-fold) partition of $\FF_q^6$ into 
$(q^2+q+1)s$ $2$-subspaces and $q^3+1 - (q+1)s$ $3$-subspaces.
\end{corollary}

\section{Characterization of infinite series of multispreads}\label{s:param}

In this section our aim is to characterize
the parameters of multispreads for small values of~$t$.
By Lemma~\ref{l:fold}, Proposition~\ref{c:ness}, and
Lemma~\ref{l:rec}(a), we have the following:
\begin{proposition}\label{p:m0}
    For $m \equiv 0 \bmod t$ and any $\mu\ge 1$,
$\param{\lambda}{\mu}{t}{m}{q}$-multispreads
exist if and only if $\lambda \equiv 0 \bmod q^t-1$.
\end{proposition}
In Section~\ref{s:t2} we completely resolve the remaining case
$m\equiv 1\bmod t$ for $t=2$.
In Section~\ref{s:t3}, for $t=3$, we close the case
$m\equiv 1\bmod t$ for arbitrary $q$ and the case $m\equiv 2\bmod t$ for $q=2,3$.
In Section~\ref{s:t4}, for $t=4$, we resolve the case
$m\equiv 2\bmod t$ for arbitrary $q$ and the cases $m\equiv 1,3\bmod t$ for $q=2$.

\subsection
{Pseudodimension \texorpdfstring{$t=2$}{t=2}}
\label{s:t2}

\begin{theorem}\label{th:t2}
Condition~\eqref{eq:mod} is necessary and sufficient for the existence of a
$\param{\lambda}{\mu}{t}{m}{q}$-multispread for $t=2$,
any $m\geq 2$, and any $\mu\geq q$.
\end{theorem}
\begin{proof}
By Corollary~\ref{c:suff},
it is sufficient to construct
$\param{\lambda}{\mu}{2}{m}{}$-multispreads 
for $m=3$, 
for all $\mu$ in  $\{q,\ldots,2q\}$,
and for the minimum~$\lambda$ such that
$$ \lambda \equiv -\mu(q^3-1) \bmod q^2-1.$$
The last relation implies that $l=\frac{\lambda}{q-1}$ is an integer
and can be rewritten as follows:
$$ \mu+l \equiv 0 \bmod q+1.$$
 So, we are to construct
 a~$\param{(q-1)l}{2q+2-l}{2}{3}{q}$-multispread for $l=2,\ldots,q$,
 a~$\para{0}{q+1}$-multispread, and a~$\para{q-1}{q}$-multispread.
 A~$\para{0}{q+1}$-multispread is solved as a special case in Section~\ref{s:special};
 a~$\para{q-1}{q}$-multispread exists by Lemmas~\ref{l:rec}(d) and~\ref{l:fold}.

In order to construct the remaining $q-1$ multispreads, we use the following recursive approach.
We start with
the $\para{0}{2q+2}$-multispread~$\bar S$
of all $2$-subspaces of~$\FF_q^3$  taken
with multiplicity~$2$.\newcommand{\pp}[1]{{\dot{#1}}}
Then, we choose
$q+1$ $1$-subspaces
$T_1$,
\ldots,
$T_{q+1}$
such that no three of them 
span
a $2$-subspace. Such a configuration 
is known to exist, see e.g.~\cite[Theorem 8.1.3]{Hirschfeld79}; for $q$ odd, 
it is called an oval in the projective plane PG$(2,q)$.

We set $\bar S_0:=\bar S$ and
for $l=1,\ldots,q$  recursively define $\bar S_l$ by taking the $q+1$ different $2$-subspaces in $\bar S_{l-1}$ incident to $T_l$ and replacing them with~$T_l$.
By the choice of $T_1$, \ldots, $T_l$,
each $2$-subspace includes at most two of them, and the initial multiplicity~$2$
of it in~$\bar S$ is sufficient for all
such replacements.
By Lemma~\ref{l:t-1}, $\bar S_l$ is a desired $\param{(q-1)l}{2q+2-l}{2}{3}{}$-multispread for $l=2,\ldots,q$.
\end{proof}


\subsection
{Pseudodimension \texorpdfstring{$t=3$}{t=3}}
\label{s:t3}
Due to Corollary \ref{c:suff}, we are to tackle the cases
$m\equiv 1 \bmod 3$ and
$m\equiv 2 \bmod 3$.
While the first case is completely solved,
see Theorem~\ref{th:t3m4} below,
the second one remains open in general,
and in Section~\ref{s:t3m5}
we discuss the solution for $q=2$ and $q=3$
and the first open questions for general~$q$.

\subsubsection{\texorpdfstring{$t=3$, $m\equiv 1\bmod 3$}{t=3, m=1 mod 3}}\label{s:t3m4}
\begin{theorem}\label{th:t3m4}
Equality~\eqref{eq:mod} is necessary and sufficient for the existence of a $\param{\lambda}{\mu}{3}{m}{q}$-multispread for any $m\equiv 1 \bmod 3$, $m\geq 4$, and any $\mu\geq q$.
\end{theorem}
\begin{proof}
By Corollary~\ref{c:suff},
it is sufficient to show the existence of a $\param{\lambda_{\mu}}{\mu}{3}{4}{q}$-multispread for all
$\mu$ in $\{q,\ldots,q^2+2q\}$ and
some 
$ \lambda_{\mu}$
such that
$\lambda_{\mu}<q^3-1$. It will immediately follow that $ \lambda_{\mu}$ is~$\lambda_{\min}$, in the notation of Corollary~\ref{c:suff}).

Consider a spread of $\FF_q^4$ into $q^2+1$ $2$-subspaces.
According to Lemma~\ref{l:rec}(e), it can be treated as a
$\param{(q^2+1)(q-1)}{q}{3}{4}{q}$-multispread.
With Lemma~\ref{l:t+1}, we obtain
$\param{\lambda_\mu}{\mu}{3}{4}{q}$-multispreads,
$\lambda_\mu=(q^2+1-i)(q-1)<q^3-1$, $\mu=q+i$
, for all
$i\in \{0,\ldots,q^2+1\}$.

The last $q-1$ values of $\mu$ are solved with Lemma~\ref{l:sum}
as follows. For every $j\in \{1,\ldots,q-1\}$,
a  $\param{\lambda_{\mu}}{\mu}{3}{4}{q}$-multispread
with $\mu=q^2+q+1+j$
can be obtained as the union of the $\param{(q^2+1-j)(q-1)}{q+j}{3}{4}{q}$- and
$\param{q(q-1)}{q^2+1}{3}{4}{q}$-multispreads constructed above.
\end{proof}

\subsubsection{\texorpdfstring{$t=3$, $m\equiv 2\bmod 3$}{t=3, m=2 mod 3}}\label{s:t3m5}

To solve the case $t=3$, $m\equiv 2\bmod 3$ for $q=2,3$,
the following multispreads have been found computationally, using integer linear programming (ILP) solvers in SAGE~\cite{sage} and GAMS~\cite{Gams}:
\begin{itemize}
 \item a $\param{5}{3}{3}{5}2$-multispread, consisting of $5$ subspaces of dimension~$2$ and $9$ subspaces of dimension~$3$;
 \item a $\param{20}{4}{3}{5}3$-multispread, consisting of $10$ subspaces of dimension~$2$ and $28$ subspaces of dimension~$3$;
 \item a $\param{12}{5}{3}{5}3$-multispread, consisting of $6$ subspaces of dimension~$2$ and $41$ subspaces of dimension~$3$.
\end{itemize}
Examples of such multispreads can be found in \ref{Appendix:A1}.

\begin{theorem}\label{th:t=3m=4}
A $\param{\lambda}{\mu}{3}{m}{q}$-multispread
exists for~$q=2,3$,
any $m\equiv 2\bmod 3$, $m\ge 5$, any $\mu\ge q$, and any $\lambda$ satisfying~\eqref{eq:mod}.
\end{theorem}

\begin{proof}
 By Corollary~\ref{c:suff},
 it is sufficient to prove the claim for
 $\mu\in \{q,\ldots,q+(q^3-1)/(q-1)-1\}$
 and
 $\lambda=\lambda_{\min}$.

 For $q=2$, the corresponding values $(\lambda_{\min},\mu)$
 are $(1,2)$, $(5,3)$, $(2,4)$, $(6,5)$, $(3,6)$, $(0,7)$, $(4,8)$. A $\para{1}{2}$-multispread is constructed as the projection (Lemma~\ref{l:rec}(d)) of a spread of $2$-subspaces of~$\FF_2^6$;
 a $\para{5}{3}$-multispread was found computationally;
 a $(0,7)$-multispread is a multifold spread (Lemma~\ref{l:fold});
 the other cases are solved with Lemma~\ref{l:sum}:
 $(2,4)=(1,2)+(1,2)$,
 $(6,5)=(1,2)+(5,3)$,
 $(3,6)=(1,2)+(2,4)$,
 $(4,8)=(1,2)+(3,6)$.

 For $q=3$, the corresponding values $(\lambda_{\min},\mu)$
 are the following:
$ (2, 3) $, $(20, 4) $, $(12, 5) $, $(4, 6)  $, $(22, 7) $, $(14, 8) $, $(6, 9)  $, $(24, 10)$, $(16, 11)$, $(8, 12) $, $(0, 13) $, $(18, 14)$, $(10, 15)$.
 A $\para{2}{3}$-multispread is constructed as the projection (Lemma~\ref{l:rec}(d)) of a spread;
 a $\para{20}{4}$- and $\para{12}{5}$-multispreads are found computationally;
 a $(0,13)$-multispread is a multifold spread (Lemma~\ref{l:fold});
 the other cases are solved with Lemma~\ref{l:sum}.
\end{proof}

For $q=4$, it is sufficient to solve
the three cases $(51,5)$, $(36,6)$, $(21,7)$ to complete the entire sequence. In general, a $\param{q-1}{q}{3}{5}{q}$-multispread
can be constructed as the projection of a $\param{0}{1}{3}{6}{q}$-multispread (spread); the parameters for
the next three cases are shown in Table~\ref{t:t3m5}.
Note that for $\mu<q^2$, only $t$- and $(t-1)$-subspaces
can occur in a $\param{\lambda_{\min}}{\mu}{t}{m}{q}$-multispread,
and hence the number of such subspaces is uniquely determined
by the other parameters. For example, 
the existence problem 
for $\param{\lambda_{\min}}{q+1}{3}{5}{q}$-multispreads
(see Table~\ref{t:t3m5} for small parameters)
can be generalized to the parameters 
$\param{\lambda_{\min}}{q+1}{s+1}{2s+1}{q}$ 
and formulated as follows:
\begin{problem}\label{pro:q2q3}
 For $s\ge 2$,
 does there exist a collection $S$
 of $q^s+1$ $s$-subspaces
 and $q^{s+1}+1$ $(s+1)$-subspaces of~$\FF_q^{2s+1}$ such that every nonzero vector
 belongs to exactly one $s$-subspace
 and one $(s+1)$-subspace from $S$
 or only to $q+1$ $(s+1)$-subspaces from~$S$?
 Equivalently (via Theorem~\ref{th:dual}),
 does there exist a $2$-fold partition of~$\FF_q^{2s+1}$
 into $q^{s+1}+1$ $s$-subspaces
 and $q^{s}+1$ $(s+1)$-subspaces?
 There are computational solutions for $s=2$, $q=2,3$ and for $s=3$, $q=2$,
 see \ref{Appendix:A1}.
\end{problem}

\begin{table}[ht]
$$
\begin{tabular}{|c|cc|cc|l}
\multicolumn{3}{|c|}{parameters ($t=3$, $m=5$, $\nu=2$)} & \multicolumn{2}{c|}{number of subspaces} & existence \\
\hline
$q$ & $\lambda$ & $\mu$
& $2$-dim & $3$-dim
&
\\\hline\hline
$q$ & $(q^2+1)(q-1)$&$q{+}1$
& $q^2+1$ & $q^3+1$ & known for $q=2,3$
\\\hline
$2$ & $5$ & $3$ & $5$ & $9$ & $\exists$,  ILP \\
$3$ & $20$ & $4$ & $10$ & $28$& $\exists$, ILP  \\
$4$ & $51$ & $5$ & $17$ & $65$ & ? \\
$5$ & $104$ & $6$ & $26$ & $126$ & ? \\
\hline
$q$ & $(q^2{-}q)(q{-}1)$&$q{+}2$
& $q^2{-}q$ & $q^3{+}q^2{+}q{+}2$ & known for $q=2,3$
\\\hline
$2$ & $2$ & $4$ & $2$ & $16$ & $\exists$, $(1,2)+(1,2)$ \\
$3$ & $12$ & $5$ & $6$ & $41$ & $\exists$, ILP  \\
$4$ & $36$ & $6$ & $12$ & $86$ & ? \\
$5$ & $80$ & $7$ & $20$ & $157$ & ? \\
\hline
$q$ & $(q^2{-}2q{-}1)(q{-}1)$&$q{+}3$
& $q^2{-}2q{-}1$ & $q^3 {+} 2q^2 {+} 2q {+} 3 $ & known for $q=3$
\\\hline
$3$ & $4$ & $6$ & $2$ & $54$ & $\exists$, $(2,3)+(2,3)$\\
$4$ & $21$ & $7$ & $7$ & $107$ & ? \\
$5$ & $56$ & $8$ & $14$ & $188$ & ? \\
\hline
\end{tabular}
$$
 \caption{First multispread parameters for $t=3$, $m=5$}\label{t:t3m5}
\end{table}


\subsection{Pseudodimension \texorpdfstring{$t=4$}{t=4}}
\label{s:t4}
For $t=4$, we start with the case $m\equiv 2\bmod 4$,
which is solved for any $q$ with help of the construction
in Section~\ref{s:desarg}.
Then, for $q=2$ only, we consider the cases $m\equiv 3\bmod 4$
and  $m\equiv 1\bmod 4$.

\subsubsection{\texorpdfstring{$t=4$, $m\equiv 2\bmod 4$}{t=4, m=2 mod 4}}\label{s:t4m6}

\begin{theorem}\label{th:t=4m=6}
A $\param{\lambda}{\mu}{4}{m}{q}$-multispread
exists for any prime power~$q$,
any $m\equiv 2\bmod 4$, $m\ge6$, any $\mu\ge q$, and any $\lambda$ satisfying~\eqref{eq:mod}.
\end{theorem}

\begin{proof}
By Corollary~\ref{c:suff}, it is sufficient
to show  the existence of a $\param{\lambda_{\mu}}{\mu}{4}{6}{q}$-multispreads for all~$\mu$ in $\{q,\ldots,q^2+q\}$,
where
$\lambda_{\mu}<q^4-1$. It will immediately follow that $ \lambda_{\mu}$ is~$\lambda_{\min}$, in notation of Corollary~\ref{c:suff}.

By Theorem~\ref{th:des}, we have a
$\param{\lambda_{\mu}}{\mu}{4}{6}{q}$-multispread,
 where $\lambda_{\mu}=(q^2-1)(q^2+1-\mu)<q^4-1$, \linebreak[4]
for all $\mu$ in $\{q,\ldots,q^2+1\}$.
It remains to construct a
$\para{\lambda_{\mu}}{\mu}$-multispread for every~$\mu$ in \linebreak[4]
$\{q^2+2,\ldots,q^2+q\}$.
It can be built as the union
of a $\para{\lambda_q}{q}$-multispread
and a $\para{\lambda_{\mu-q}}{\mu-q}$-multispread,
since
$$
\lambda_q +\lambda_{\mu-q} = (q^2-1)(q^2+1+(q^2+1)-\mu) < q^4-1
$$
if $\mu>q^2+1$.
\end{proof}

\subsubsection{\texorpdfstring{$t=4$, $m\equiv 3\bmod 4$, $q=2$}{t=4, m=3 mod 4, q=2}}\label{s:t4m7}

\begin{theorem}\label{th:t=4m=7}
A $\param{\lambda}{\mu}{4}{m}{2}$-multispread
exists for
any $m\equiv 3\bmod 4$, $m\ge 7$, any $\mu\ge 2$, and any $\lambda$ satisfying~\eqref{eq:mod}.
\end{theorem}

\begin{proof}
 By Corollary~\ref{c:suff},
 it is sufficient to prove the claim for
 $\mu\in \{2,\ldots,16\}$; the corresponding pairs
 $(\lambda_{\min},\mu)$ are
$(1, 2)  $,
$(9, 3)  $,
$(2, 4)  $,
$(10, 5) $,
$(3, 6)  $,
$(11, 7) $,
$(4, 8)  $,
$(12, 9) $,
$(5, 10) $,
$(13, 11)$,
$(6, 12) $,
$(14, 13)$,
$(7, 14) $,
$(0, 15) $,
$(8, 16) $.

A $\param{1}{2}{4}{7}{2}$-multispread can be constructed as the projection
(Lemma~\ref{l:rec}(d))
of a $\param{0}{1}{4}{8}{2}$-multispread (spread).
A $\param{9}{3}{4}{7}{2}$-multispread was found
computationally, see \ref{Appendix:A1}.
A $\param{0}{15}{4}{7}{2}$-multispread
is a $15$-fold spread (Lemma~\ref{l:fold}).
The other cases are solved with Lemma~\ref{l:sum}.
\end{proof}

\subsubsection{\texorpdfstring{$t=4$, $m\equiv 1\bmod 4$, $q=2$}{t=4, m=1 mod 4, q=2}}\label{s:t4m5}

By Corollary~\ref{c:l2452}, there are no
$\param{\lambda}{2}{4}{5}{2}$- and $\param{\lambda}{3}{4}{5}{2}$-multispreads.
The remaining parameters are covered by the following theorem.

\begin{theorem}\label{th:t=4m=5}
A $\param{\lambda}{\mu}{4}{m}{2}$-multispread
exists for
any $m\equiv 1\bmod 4$, $m\ge 5$, any $\mu\ge 2$, and any $\lambda$ satisfying~\eqref{eq:mod}, except the cases when $m=5$ and $\mu\in\{2,3\}$.
\end{theorem}

\begin{proof}
The cases $\mu=2$ and $\mu=3$ are special, because the minimum~$m$ is~$9$.
Multispreads with parameters $\param{13}{2}{4}{9}{2}$ and
$\param{12}{3}{4}{9}{2}$ 
were found computationally, with predefined automorphism group~$\Sc{2^3}$, see \ref{Appendix:A2}. The larger values of $m$ and $\lambda$
are covered by Lemma~\ref{l:rec}(c,a).

 For the other values of $\mu$,
 by Corollary~\ref{c:suff},
 it is sufficient to consider $\mu$ from $\{4,\ldots,18\}$; the corresponding pairs
 $(\lambda_{\min},\mu)$ are
 $(15-\mu,\mu)$, $\mu=4,5,\ldots,15$,
 and $(14,16)$, $(13,17)$, $(12,18)$. As the last
 three pairs are solved from the first $12$ pairs with Lemma~\ref{l:sum},
 it remains to solve the pairs of form $(15-\mu,\mu)$.

 By Lemma~\ref{l:rec}(e), a $\param{1}{2}{3}{5}{2}$-multispread
 constructed as in Section~\ref{s:t3m5}
 and consisting of one $2$-subspace and eight $3$-subspaces
 is also an $\param{11}{4}{4}{5}{2}$-multispread.
 We can apply Lemma~\ref{l:t+1} to replace
 each of the $3$-subspaces by three $4$-subspaces, which increases $\mu$ by~$1$. In such a way, we obtain $\param{15-\mu}{\mu}{4}{5}{2}$-multispreads
 for $\mu=4,5,\ldots,12$.

 The set $S$ of all $4$-subspaces of $\FF_2^5$ is a
 $\param{0}{15}{4}{5}{2}$-multispread.
 There are nine $3$-subspaces of~$\FF_2^5$
 such that no two of them are included in the same $4$-subspace (see, e.g.,
 \cite{HKKW2016Tables} and the corresponding table \url{http://subspacecodes.uni-bayreuth.de/table/2/5/4/2/}).  Therefore, Lemma~\ref{l:t-1} can be applied to~$S$ up to $9$ times,
 producing $\param{15-\mu}{\mu}{4}{5}{2}$-multispreads
 for $\mu=15,14,\ldots,6$.
\end{proof}

Note that the two sequences constructed in the proof
intersect by parameters, but the corresponding multispreads
are different: each multispread from the first sequence ($\mu=4,5,\ldots,12$)
contains a $2$-subspace, while the multispreads
from the second sequence ($\mu=15,14,\ldots,6$) do not contain
$2$-subspaces.
The following observation is of independent interest;
it shows that $t$- and $(t-1)$-subspaces are not always
enough to construct a multispread with required parameters.
\begin{proposition}\label{p:dim2}
 Every multispread with parameters $\param{11}{4}{4}{5}{2}$ or $\param{10}{5}{4}{5}{2}$ contains \linebreak[4]a $2$-subspace.
\end{proposition}
\begin{proof}
Since $\mu<8$, there are no $1$-subspaces in the multispreads;
since $\lambda<15$, there are no $0$-subspaces.
Considering the number of $2$-, $3$-, and $4$-subspaces, respectively $x$, $y$, and~$z$, as indeterminates,
we have two linear equations, one for~$\lambda$ and one for~$\mu$. Solving them gives $y=11-3x$, $z=-2+2x$
for a $\param{11}{4}{4}{5}{2}$-multispread (immediately we have $x \ne 0$)
and  $y=10-3x$, $z=1+2x$
for a $\param{10}{5}{4}{5}{2}$-multispread. In the last case $x=0$
implies that there is only one $4$-subspace in the multispread, and hence only~$15$ of~$31$ nonzero vectors of the space are covered with odd multiplicity. Since $\mu=5$ is odd, this leads to a contradiction.
\end{proof}

\begin{remark}
   In the discussion above, we touched on a question that is not actually considered in the current study, the existence of multispreads  
   with given number of subspaces of each dimension (the equivalent question is for dual multifold partitions).
   Taking into account Proposition~\ref{p:dim2} and the simple argument that the dual partition is $1$-fold and has not more than one subspace of dimension~$3$ or~$4$, it is easy to see that for $\param{15-\mu}{\mu}{4}{5}{2}$-multispreads the constructions from the proof of Theorem~\ref{th:t=4m=5}
   exhaust all possibilities except one for $\param{7}{8}{4}{5}{2}$,
   with one $1$-dimensional subspace.
   Such example can be constructed by projecting (Lemma~\ref{l:rec}(d)) a spread in~$\FF_2^8$ three times. For many other parameters, the question remains unsolved.
\end{remark}

\subsection*{Acknowledgments}
The authors thank Sascha Kurz for useful discussions.
The work was funded by the Russian Science Foundation (22-11-00266),
\url{https://rscf.ru/project/22-11-00266/}.



\begin{thebibliography}{10}
\expandafter\ifx\csname url\endcsname\relax
  \def\url#1{\texttt{#1}}\fi
\expandafter\ifx\csname urlprefix\endcsname\relax\def\urlprefix{URL }\fi
\expandafter\ifx\csname href\endcsname\relax
  \def\href#1#2{#2} \def\path#1{#1}\fi

\bibitem{BMP:2024}
J.~Bierbrauer, S.~Marcugini, F.~Pambianco, An asymptotic property of quaternary
  additive codes, \href{http://link.springer.com/journal/10623}{Des. Codes
  Cryptography} 92~(11) (2024) 3371--3375, 
\newblock \href {https://doi.org/10.1007/s10623-024-01438-2}
  {\path{https://doi.org/10.1007/s10623-024-01438-2}}.

\bibitem{BMP:2021}
J.~Bierbrauer, S.~Marcugini, F.~Pambianco, Optimal additive quaternary codes of
  low dimension,
  \href{http://ieeexplore.ieee.org/xpl/RecentIssue.jsp?punumber=18}{IEEE Trans.
  Inf. Theory} 67~(8) (2021) 5116--5118, 
\newblock \href {https://doi.org/10.1109/TIT.2021.3085577}
  {\path{https://doi.org/10.1109/TIT.2021.3085577}}.

\bibitem{BlokBrow:2004}
A.~Blokhuis, A.~E. Brouwer, Small additive quaternary codes,
  \href{http://www.sciencedirect.com/science/journal/01956698}{Eur. J. Comb.}
  25~(2) (2004) 161--167, 
\newblock \href {https://doi.org/10.1016/S0195-6698(03)00096-9}
  {\path{https://doi.org/10.1016/S0195-6698(03)00096-9}}.

\bibitem{Bonisoli}
A.~Bonisoli, Every equidistant linear code is a sequence of dual {H}amming
  codes, Ars Combin. 18 (1984) 181--186.

\bibitem{Bu:1980:pertitions}
T.~Bu, Partitions of a vector space,
  \href{http://www.sciencedirect.com/science/journal/0012365X}{Discrete Math.}
  31 (1980) 79--83,  
\newblock \href {https://doi.org/10.1016/0012-365X(80)90174-0}
  {\path{https://doi.org/10.1016/0012-365X(80)90174-0}}.

\bibitem{Delsarte:1973}
P.~Delsarte, An Algebraic Approach to Association Schemes of Coding Theory,
  Vol.~10 of Philips Res. Rep., Supplement, N.V.~Philips' Gloeilampenfabrieken,
  Eindhoven, 1973.

\bibitem{El-Zanati-et-al:lambda}
S.~El-Zanati, G.~Seelinger, P.~Sissokho, L.~Spence, C.~Vanden~Eynden, On
  $\lambda$-fold partitions of finite vector spaces and duality,
  \href{http://www.sciencedirect.com/science/journal/0012365X}{Discrete Math.}
  311~(4) (2011) 307--318,  
\newblock \href {https://doi.org/10.1016/j.disc.2010.10.026}
  {\path{https://doi.org/10.1016/j.disc.2010.10.026}}.

\bibitem{Gams}
GAMS Development Corporation. General Algebraic Modeling System (GAMS) Release
  24.2.1., Washington, DC, USA, 2013.

\bibitem{Heden:2012:survey}
O.~Heden, A survey of the different types of vector space partitions,
  \href{https://www.worldscientific.com/loi/dmaa}{Discrete Math. Algorithms
  Appl.} 4~(1) (2012) 1250001(1--14),  
\newblock \href {https://doi.org/10.1142/S1793830912500012}
{\path{https://doi.org/10.1142/S1793830912500012}}.


\bibitem{HKKW2016Tables}
D.~Heinlein, M.~Kiermaier, S.~Kurz, A.~Wassermann,
   {Tables of subspace codes}, E-print
  1601.02864, arXiv.org  (2016),
\newblock \href {https://doi.org/10.48550/arXiv.1601.02864}
  {\path{https://doi.org/10.48550/arXiv.1601.02864}}.

\bibitem{Hirschfeld79}
J.~W.~P. Hirschfeld, Projective Geometries Over Finite Fields, Oxford
  University Press, New York, 1979.

\bibitem{KasPot08}
P.~Kaski, O.~Pottonen,
  \href{https://researchportal.helsinki.fi/en/publications/libexact-users-guide-version-10}{libexact
  user's guide, version 1.0}, Tech. Rep. 2008-1, Helsinki Institute for
  Information Technology HIIT (2008).

\bibitem{Kro:mufold}
D.~S. Krotov, Multifold $1$-perfect codes,
  \href{http://onlinelibrary.wiley.com/journal/10.1002/(ISSN)1520-6610}{J.
  Comb. Des.} 32~(9) (2024) 546--555, 
\newblock \href {https://doi.org/10.1002/jcd.21947}
  {\path{https://doi.org/10.1002/jcd.21947}}.

\bibitem{KroPot:Ch1}
D.~S. Krotov, V.~N. Potapov, Completely regular codes and equitable partitions,
  in: 
M.~Shi, P.~Sol\'e (Eds.), Completely Regular Codes in Distance Regular Graphs,
 CRC Press, Boca Raton FL, 2025, Ch.~1, pp. 1--84, 
\newblock \href {https://doi.org/10.1201/9781003393931-1}
  {\path{https://doi.org/10.1201/9781003393931-1}}.

\bibitem{MatTop:2009}
Z.~T. Mateva, S.~T. Topalova, Line spreads of {PG(5,2)},
  \href{http://onlinelibrary.wiley.com/journal/10.1002/(ISSN)1520-6610}{J.
  Comb. Des.} 17~(1) (2009) 90--102, 
\newblock \href {https://doi.org/10.1002/jcd.20198}
  {\path{https://doi.org/10.1002/jcd.20198}}.

\bibitem{sage}
{The Sage Developers}, {S}ageMath, the {S}age {M}athematics {S}oftware {S}ystem
  ({V}ersion 10.0), {\tt https://www.sagemath.org} (2023).

\bibitem{Zinoviev:Ch2}
V.~A. Zinoviev, Completely regular codes over finite fields, in: 
M.~Shi, P.~Sol\'e (Eds.), Completely Regular Codes in Distance Regular Graphs,
 CRC Press, Boca Raton FL, 2025, Ch.~2, pp. 85--205,  
\newblock \href {https://doi.org/10.1201/9781003393931-2}
  {\path{https://doi.org/10.1201/9781003393931-2}}.

\end{thebibliography}

\providecommand\href[2]{#2} \providecommand\url[1]{\href{#1}{#1}}
  \def\DOI#1{{\href{https://doi.org/#1}{https://doi.org/#1}}}\def\DOIURL#1#2{{\href{https://doi.org/#2}{https://doi.org/#1}}}

\appendix
\section{Multispreads found by computation}\label{Appendix}
Below, we provide some examples of multispreads
(namely,
$\param{5}{3}{3}{5}{2}$-,
\mbox{$\param{20}{4}{3}{5}{3}$-,}
\mbox{$\param{12}{5}{3}{5}{3}$-,}
\mbox{$\param{9}{3}{4}{7}{2}$-,}
\mbox{$\param{13}{2}{4}{9}{2}$-,} and
$\param{12}{3}{4}{9}{2}$-multispreads)
found using ILP solvers (from GAMS~\cite{Gams}, SAGE~\cite{sage}, and for $\param{5}{3}{3}{5}{2}$, by libexact~\cite{KasPot08}). In the last two cases, for $m=9$,
the search was based on the predefined automorphism group~$\Sc{2^3}$.

\subsection{Solutions of Problem~\ref{pro:q2q3} for particular cases}\label{Appendix:A1}
By Theorem~\ref{th:dual}, the multispreads in this section correspond to 
$2$-fold partitions of the space, via duality.

The \underline{$\param{5}{3}{3}{5}{2}$-multispreads} were classified computationally,
using \texttt{libexact}~\cite{KasPot08} for exhaustive search.
There are 88 equivalence classes, with automorphism group orders $1$, $2$, $3$, $4$, $6$, and $12$. Representatives of the four classes of the most symmetric multispreads are listed below. Their common automorphism group, with the structure $S_3 \times C_2$, is generated by
$$
\left(\begin{array}{cc@{\ }cc@{\ }c}
1&0&0&0&0\\
0&0&1&0&0\\[-0.2em]  
0&1&1&0&0\\   
0&0&0&0&1\\[-0.2em] 
0&0&0&1&1
\end{array}\right)
,\quad
\left(\begin{array}{@{\ }cc@{\ }cc@{\ }c@{\ }}
1&0&0&0&0\\
0&0&1&0&0\\[-0.2em]
0&1&0&0&0\\   
0&0&0&0&1\\[-0.2em]   
0&0&0&1&0
\end{array} \right)                 
,\quad
\left(\begin{array}{cc@{\ }cc@{\ }c}
1&0&0&0&0\\
0&0&0&1&0\\[-0.2em]                 
0&0&0&0&1\\                 
0&1&0&0&0\\[-0.2em]                 
0&0&1&0&0
\end{array}\right).$$
By $\Orb\, x$, we denote the orbit of a subspace $x$ under this automorphism group. The four multispreads $X_1$, $X_2$, $X_3$, $X_4$ are defined as follows:

$A = \Orb\langle 0\,00\,01,  0\,00\,10 \rangle\cup \Orb\langle 1\,01\,10,  1\,10\,01 \rangle,$

$ B=\Orb\langle 0\,01\,11, 0\,11\,10, 1\,11\,11 \rangle,\quad B'=\Orb\langle 1\,00\,01, 1\,00\,10, 1\,11\,11 \rangle,$

$ X_1 = \Orb \langle 0\,00\,01,  0\,00\,10,  1\,00\,00 \rangle \cup \Orb \langle 0\,01\,01,  0\,10\,10,  1\,00\,00 \rangle \cup B \cup A,$

$ X_2 = \Orb \langle 0\,00\,01,  0\,01\,00,  1\,00\,00 \rangle \cup B \cup A,$

$X_3 = \Orb \langle 1\,01\,10,  1\,10\,01,  1\,00\,00 \rangle \cup B' \cup A,$

$X_4 = 
\Orb \langle 0\,01\,11,  0\,11\,10,  1\,00\,00 \rangle 
\cup
\Orb \langle 0\,01\,01,  0\,10\,10,  1\,00\,00 \rangle 
\cup B' \cup A.$

A \underline{$\param{20}{4}{3}{5}{3}$-multispread}:\\
 $\{ \langle 00101,00012 \rangle$,
 $\langle 10220,01002 \rangle$,
 $\langle 10011,01010 \rangle$,
 $\langle 12010,00120 \rangle$,
 $\langle 10120,01211 \rangle$,
 \linebreak[4]
 $\langle 10021,01200 \rangle$,
 $\langle 10020,01210 \rangle$,
 $\langle 10210,01021 \rangle$,
 $\langle 01000,00111 \rangle$,
 $\langle 10012,00112 \rangle$,
 \linebreak[4]
 $\langle 10021,01022,00120 \rangle$,
 $\langle 10002,01020,00111 \rangle$,
 $\langle 10102,01100,00010 \rangle$,
 $\langle 10002,01012,00121 \rangle$,
 \linebreak[4]
 $\langle 10022,01001,00121 \rangle$,
 $\langle 10100,01001,00011 \rangle$,
 $\langle 10022,01011,00121 \rangle$,
 $\langle 10000,01010,00001 \rangle$,
 \linebreak[4]
 $\langle 11100,00010,00001 \rangle$,
 $\langle 10100,01200,00001 \rangle$,
 $\langle 10010,01002,00101 \rangle$,
 $\langle 10202,01101,00011 \rangle$,
 \linebreak[4]
 $\langle 10010,01012,00121 \rangle$,
 $\langle 10000,01102,00010 \rangle$,
 $\langle 10022,01020,00102 \rangle$,
 $\langle 10001,01022,00102 \rangle$,
 \linebreak[4]
 $\langle 10002,01021,00100 \rangle$,
 $\langle 10011,01011,00102 \rangle$,
 $\langle 10012,01001,00100 \rangle$,
 $\langle 10202,01202,00010 \rangle$,
 \linebreak[4]
 $\langle 10022,01020,00100 \rangle$,
 $\langle 10001,01001,00100 \rangle$,
 $\langle 10100,01110,00001 \rangle$,
 $\langle 10000,01011,00112 \rangle$,
 \linebreak[4]
 $\langle 10200,01102,00011 \rangle$,
 $\langle 10000,01000,00122 \rangle$,
 $\langle 10010,01022,00102 \rangle$,
 $\langle 10001,01022,00110 \rangle \}.$

A \underline{$\param{12}{5}{3}{5}{3}$-multispread}:\\
$\{ \langle 10112,01210\rangle$,
 $\langle 01011,00111\rangle$,
 $\langle 10201,01111\rangle$,
 $\langle 12010,00001\rangle$,
 $\langle 10000,01110\rangle$,
 \linebreak[4]
 $\langle 11000,00102\rangle$,
 $\langle 10100,01002,00010\rangle$,
 $\langle 10202,01001,00011\rangle$,
 $\langle 10010,00120,00001\rangle$,
 \linebreak[4]
 $\langle 10202,01002,00011\rangle$,
 $\langle 10011,01011,00120\rangle$,
 $\langle 10021,01022,00110\rangle$,
 $\langle 10001,01202,00010\rangle$,
 \linebreak[4]
 $\langle 10002,00100,00012\rangle$,
 $\langle 10010,01021,00110\rangle$,
 $\langle 10020,01000,00110\rangle$,
 $\langle 10002,01000,00111\rangle$,
 \linebreak[4]
 $\langle 10002,01002,00011\rangle$,
 $\langle 10020,01022,00120\rangle$,
 $\langle 10022,01001,00111\rangle$,
 $\langle 10001,01000,00121\rangle$,
 \linebreak[4]
 $\langle 10022,01012,00121\rangle$,
 $\langle 10021,01022,00101\rangle$,
 $\langle 10010,01000,00101\rangle$,
 $\langle 10010,01020,00101\rangle$,
 \linebreak[4]
 $\langle 10102,01001,00012\rangle$,
 $\langle 01100,00010,00001\rangle$,
 $\langle 10001,01202,00011\rangle$,
 $\langle 10002,01000,00102\rangle$,
 \linebreak[4]
 $\langle 10020,01001,00101\rangle$,
 $\langle 10000,01002,00121\rangle$,
 $\langle 11001,00100,00010\rangle$,
 $\langle 10022,01021,00102\rangle$,
 \linebreak[4]
 $\langle 10012,01012,00112\rangle$,
 $\langle 10012,01020,00122\rangle$,
 $\langle 10002,01020,00100\rangle$,
 $\langle 10001,01010,00112\rangle$,
 \linebreak[4]
 $\langle 10201,01202,00012\rangle$,
 $\langle 12000,00101,00012\rangle$,
 $\langle 10022,01002,00122\rangle$,
 $\langle 10011,01021,00112\rangle$,
 \linebreak[4]
 $\langle 10012,01012,00100\rangle$,
 $\langle 10102,01201,00010\rangle$,
 $\langle 10022,01001,00100\rangle$,
 $\langle 10021,01020,00112\rangle$,
 \linebreak[4]
 $\langle 10000,01011,00122\rangle$,
 $\langle 10001,01010,00120\rangle \}$.

A \underline{$\param{9}{3}{4}{7}{2}$-multispread} (in the hexadecimal form):
$\{
\langle 40,20,10\rangle $,
$\langle 41,22,14\rangle $,
$\langle 42,2d,1c\rangle $,
$\langle 45,2c,13\rangle $,
$\langle 49,2e,12\rangle $,
$\langle 52,24,0e\rangle $,
$\langle 62,16,09\rangle $,
$\langle 65,08,03\rangle $,
$\langle 23,19,04\rangle $,
\linebreak[4]
$\langle 42,25,15,0d\rangle $,
$\langle 43,21,13,08\rangle $,
$\langle 43,27,12,0c\rangle $,
$\langle 44,25,16,0e\rangle $,
$\langle 46,25,11,0a\rangle $,
$\langle 47,23,14,0c\rangle $,
$\langle 40,21,1a,07\rangle $,
$\langle 48,20,11,06\rangle $,
$\langle 48,22,11,06\rangle $,
$\langle 4a,21,19,07\rangle $,
$\langle 4c,28,1a,01\rangle $,
$\langle 43,32,0a,05\rangle $,
$\langle 45,34,0d,02\rangle $,
$\langle 54,34,0c,01\rangle $,
$\langle 41,10,0a,05\rangle $,
$\langle 44,15,09,02\rangle $,
$\langle 28,04,02,01\rangle
\}$.

\subsection{\texorpdfstring{$t=4$, $m=9$}{t=4, m=9}}
\label{Appendix:A2}

We exploit the ideas of Section~\ref{s:desarg} 
in the ILP 
search for this case. 

We first show, of independent interest,
a \ul{partition $S$ of $\FF_{2}^9$
into $28$ $3$-subspaces and $21$ $4$-subspaces}.
By repeating $4$-subspaces from~$S$ twice, we get 
a $\param{28}{2}{4}{9}{2}$-multispread.
As a $9$-dimensional space over $\FF_2$,
we take $\FF_{2^9}=\FF_2[z]/\langle P(z) \rangle $, where $P(z)=z^9 + z^4 + 1$,
and denote by~$\alpha$
a~primitive root of~$P(z)$.
The $28$ $3$-subspaces in~$S$ are
\begin{multline*}
\{0\}\cup\{\alpha^{i+73j}:\  j\in\{0,1,2,3,4,5,6\}\},\\\mbox{where }
    i\in \{0, 1, 4, 10, 11, 14,19,  21, 22, 23, 24, 25, 26, 30, 32, \\ 37, 39, 44, 49, 50, 52, 53, 55, 61, 62, 63, 70, 72 \}.
 \end{multline*}
The $21$ $4$-subspaces in $S$
are 
\begin{equation}\label{eq:bV}
   \beta V,\quad \mbox{where }\beta \in \FF_{2^3}\backslash\{0\}=\{\alpha^{73j}:\ j\in\{0,1,2,3,4,5,6\}\}
\end{equation}
and $V$ is one of the following three $4$-subspaces over~$\FF_2$:
$$
\langle \alpha^{59}, \alpha^{184}, \alpha^{363}, \alpha^{378}\rangle, \quad
\langle \alpha^{3}, \alpha^{81}, \alpha^{235}, \alpha^{332}\rangle, \quad
\langle \alpha^{36}, \alpha^{64}, \alpha^{307}, \alpha^{361}\rangle.$$
By exhaustive search, we did not find a similar construction of a partition of~$\FF_{2}^9$ into $13$ \mbox{$3$-}subspaces and $28$ $4$-subspaces without restrictions on the automorphism group, the existence of such a partition remains an open question. So we had to make a separate search for a $\param{13}{2}{4}{9}{2}$-multispread.

A \underline{$\param{13}{2}{4}{9}{2}$-multispread}
(the dual is a $5$-fold partition of $\FF_2^9$)
consists of (keeping the notation above)
\begin{itemize}
    \item 
the $13$ $3$-subspaces 
\begin{equation*}
\{0\}\cup\{\alpha^{i+73j}:\  j\in\{0,\ldots,6\}\}, \qquad
    i\in \{8,11,19,20,23,34,35,37,43,44,50,51,62\},
\end{equation*}
    \item 
and the $7\cdot 8$ $4$-subspaces of form~\eqref{eq:bV} with $V$ from
\begin{multline*}
\langle \alpha^{0}, \alpha^{1}, \alpha^{2}, \alpha^{3} \rangle, \quad
\langle \alpha^{7}, \alpha^{9}, \alpha^{453}, \alpha^{167} \rangle, \quad
\langle \alpha^{12}, \alpha^{159}, \alpha^{89}, \alpha^{178} \rangle, \quad
\langle \alpha^{0}, \alpha^{78}, \alpha^{448}, \alpha^{158} \rangle, \\
\langle \alpha^{3}, \alpha^{442}, \alpha^{225}, \alpha^{86} \rangle, \ \ 
\langle \alpha^{2}, \alpha^{372}, \alpha^{17}, \alpha^{164} \rangle, \ \ 
\langle \alpha^{5}, \alpha^{83}, \alpha^{453}, \alpha^{382} \rangle, \ \ 
\langle \alpha^{1}, \alpha^{150}, \alpha^{444}, \alpha^{374} \rangle.
\end{multline*}
\end{itemize}
\medskip

A \underline{$\param{12}{3}{4}{9}{2}$-multispread}
(the dual is a $7$-fold partition of $\FF_2^9$)
consists of  
\begin{itemize}
    \item 
the $12$ $3$-subspaces 
\begin{equation*}
\{0\}\cup\{\alpha^{i+73j}:\  j\in\{0,\ldots,6\}\},\qquad
    i \in  \{0,2,4,25,32,36,50,56,62,66,68,72\},
\end{equation*}
    \item and the $7\cdot 13$ $4$-subspaces of form~\eqref{eq:bV} with $V$ from
\begin{equation*}
\begin{array}{lll} 
     \langle \alpha^{12}, \alpha^{13}, \alpha^{191}, \alpha^{346} \rangle,&
     \langle \alpha^{27}, \alpha^{63}, \alpha^{64}, \alpha^{326} \rangle, &
     \langle \alpha^{20}, \alpha^{25}, \alpha^{26}, \alpha^{287} \rangle, \\
     \langle \alpha^{54}, \alpha^{55}, \alpha^{96}, \alpha^{487} \rangle, &
     \langle \alpha^{15}, \alpha^{16}, \alpha^{186}, \alpha^{423} \rangle, &
     \langle \alpha^{27}, \alpha^{28}, \alpha^{99}, \alpha^{394} \rangle, \\
     \langle \alpha^{2}, \alpha^{48},  \alpha^{59}, \alpha^{378} \rangle, &     \langle \alpha^{24}, \alpha^{121}, \alpha^{226}, \alpha^{354} \rangle, &
     \langle \alpha^{63}, \alpha^{85}, \alpha^{179}, \alpha^{237} \rangle, \\
     \langle \alpha^{31}, \alpha^{41}, \alpha^{97}, \alpha^{362} \rangle, &
     \langle \alpha^{59}, \alpha^{85}, \alpha^{327}, \alpha^{482} \rangle, &
     \langle \alpha^{21}, \alpha^{50}, \alpha^{104}, \alpha^{477} \rangle, \\
     \langle \alpha^{18}, \alpha^{30}, \alpha^{93}, \alpha^{280} \rangle.
\end{array}
\end{equation*}
\end{itemize}
\end{document}